\documentclass[aps,prb,reprint,preprintnumbers,superscriptaddress,amsmath,amssymb,bibnotes,longbibliography]{revtex4-2}

\usepackage{graphicx}
\usepackage{dcolumn}
\usepackage{bm}
\usepackage[colorlinks,linkcolor=blue,anchorcolor=blue,citecolor=blue,urlcolor=blue,filecolor=blue,menucolor=blue,runcolor=blue]{hyperref}

\newcommand{\beq}{\begin{equation}}
\newcommand{\eeq}{\end{equation}}
\newcommand{\bea}{\begin{eqnarray}}
\newcommand{\eea}{\end{eqnarray}}

\newcommand{\fig}[1]{Fig.~\ref{#1}}

\begin{document}
\title{Muon-spin relaxation study of the layered kagome superconductor CsV$_3$Sb$_5$}

\author{Zhaoyang Shan}
\affiliation{Center for Correlated Matter and Department of Physics, Zhejiang University, Hangzhou 310058, China}
\affiliation  {Zhejiang Province Key Laboratory of Quantum Technology and Device, Department of Physics, Zhejiang University, Hangzhou 310058, China}

\author{Pabitra K. Biswas}
\email{pabitra.biswas@stfc.ac.uk}
\affiliation{ISIS Facility, STFC Rutherford Appleton Laboratory, Harwell Science and Innovation Campus, Oxfordshire, OX11 0QX, United Kingdom}

\author{Sudeep K. Ghosh}
\email{S.Ghosh@kent.ac.uk}
\affiliation{School of Physical Sciences, University of Kent, Canterbury CT2 7NH, United Kingdom}

\author{T. Tula}
\affiliation{School of Physical Sciences, University of Kent, Canterbury CT2 7NH, United Kingdom}

\author{Adrian D. Hillier}
\affiliation{ISIS Facility, STFC Rutherford Appleton Laboratory, Harwell Science and Innovation Campus, Oxfordshire, OX11 0QX, United Kingdom}

\author{Devashibhai Adroja}
\affiliation{ISIS Facility, STFC Rutherford Appleton Laboratory, Harwell Science and Innovation Campus, Oxfordshire, OX11 0QX, United Kingdom}
\affiliation{Highly Correlated Matter Research Group, Physics Department, University of Johannesburg, P.O. Box 524, Auckland Park 2006, South Africa}

\author{Stephen Cottrell}
\affiliation{ISIS Facility, STFC Rutherford Appleton Laboratory, Harwell Science and Innovation Campus, Oxfordshire, OX11 0QX, United Kingdom}

\author{Guang-Han Cao}
\affiliation{Center for Correlated Matter and Department of Physics, Zhejiang University, Hangzhou 310058, China}
\affiliation  {Zhejiang Province Key Laboratory of Quantum Technology and Device, Department of Physics, Zhejiang University, Hangzhou 310058, China}
\affiliation  {State Key Laboratory of Silicon Materials, Zhejiang University, Hangzhou 310058, China}

\author{Yi Liu}
\affiliation  {Key Laboratory of Quantum Precision Measurement of Zhejiang Province, Department of Applied Physics, Zhejiang University of Technology, Hangzhou 310023, China}

\author{Xiaofeng Xu}
\affiliation  {Key Laboratory of Quantum Precision Measurement of Zhejiang Province, Department of Applied Physics, Zhejiang University of Technology, Hangzhou 310023, China}

\author{Yu Song}
\affiliation{Center for Correlated Matter and Department of Physics, Zhejiang University, Hangzhou 310058, China}
\affiliation  {Zhejiang Province Key Laboratory of Quantum Technology and Device, Department of Physics, Zhejiang University, Hangzhou 310058, China}

\author{Huiqiu Yuan}
\affiliation{Center for Correlated Matter and Department of Physics, Zhejiang University, Hangzhou 310058, China}
\affiliation  {Zhejiang Province Key Laboratory of Quantum Technology and Device, Department of Physics, Zhejiang University, Hangzhou 310058, China}
\affiliation  {State Key Laboratory of Silicon Materials, Zhejiang University, Hangzhou 310058, China}

\author{Michael Smidman}
\email{msmidman@zju.edu.cn}
\affiliation{Center for Correlated Matter and Department of Physics, Zhejiang University, Hangzhou 310058, China}
\affiliation  {Zhejiang Province Key Laboratory of Quantum Technology and Device, Department of Physics, Zhejiang University, Hangzhou 310058, China}
\date{\today}
\addcontentsline{toc}{chapter}{Abstract}

\begin{abstract}
The ${\mathbb{Z}}_{2}$ topological metals $R$V$_3$Sb$_5$ ($R$~=~K, Rb, Cs) with a layered kagome structure provide a unique opportunity to investigate the interplay between charge order, superconductivity and topology. Here, we report muon-spin relaxation/rotation ($\mu$SR) measurements performed on CsV$_3$Sb$_5$ across a broad temperature range, in order to uncover the nature of the charge-density wave order and superconductivity in this material. From zero-field $\mu$SR, we find that spontaneous magnetic fields appear below 50~K which is well below the charge-density wave transition ($T^* \sim 93$~K). We show that these spontaneous fields are dynamic in nature making it difficult to associate them with a hidden static order. The superconducting state of CsV$_3$Sb$_5$ is found to preserve time-reversal symmetry and the transverse-field $\mu$SR results are consistent with a superconducting state that has two fully open gaps.

\end{abstract}

\maketitle

\section{Introduction}
Kagome lattice compounds have served as ideal platforms for exploring both unusual magnetic phenomena, such as geometric frustration and quantum spin-liquids \cite{Balents2010,Mingxuan2015,Han2012}, as well as electronic behaviors due to the presence of flat bands, Dirac cones and non-trivial band topologies \cite{Lin2018,Yin2019,Kang2019,Guo2009}. The recently discovered  superconductors  $R$V$_3$Sb$_5$ ($R$~=~K, Rb, Cs) have therefore attracted much attention for the study of superconductivity in   kagome lattice systems \cite{Ortiz2019,Ortiz2020,Ortiz2021,yin2021superconductivity}, which have topological band structures with multiple Dirac cones.  These materials also exhibit  unusual charge density wave (CDW) ordering, which is accompanied by a giant anomalous Hall effect \cite{Yang2020,yu2021concurrence}, and in KV$_3$Sb$_5$ was reported to correspond to a chiral charge ordering \cite{jiang2020discovery}. Moreover, this CDW order exhibits clear competition with the superconductivity in these compounds, where there is an enhancement of $T_{\rm c}$ upon the suppression of the CDW state by pressure \cite{chen2021double,Du2021,Zhang2021,Chen2021,Wang2021b,Wang2021c,Yu2021,Du2021b}. Although low temperature thermal conductivity measurements suggested a nodal superconducting gap in CsV$_3$Sb$_5$ \cite{zhao2021nodal}, penetration depth measurements using the tunnel-diode oscillator and muon-spin rotation methods \cite{duan2021nodeless,gupta2021microscopic}, as well as scanning-tunneling microscopy, point contact spectroscopy, and nuclear magnetic/quadrupole resonance \cite{Yin2021,Liang2021,Xu2021,Mu2021,song2021orbital} point to fully gapped multiband superconductivity with a sign-preserving order parameter.

Given the unconventional properties of the CDW state, together with various theoretical proposals including  chiral flux phases, star of David and inverse star of David configurations \cite{Denner2021,Feng2021,Tan2021,Park2021}, it is important to probe whether time-reversal symmetry (TRS) is broken in the CDW state.  In the case of KV$_3$Sb$_5$, an enhanced relaxation rate of the asymmetry from muon-spin relaxation ($\mu$SR) is detected below the CDW transition $T^*=80$~K \cite{Mielke2021TRS}, corresponding to the spontaneous appearance of static magnetic fields. On the other hand, in CsV$_3$Sb$_5$ the enhanced $\mu$SR relaxation was only found to onset well below the CDW transition $T^*=93$~K \cite{yu2021evidence}, while no evidence for TRS breaking along the $c$~axis could be detected with spin-polarized tunneling spectroscopy \cite{Li2022}. As such, it is of particular interest to further examine the nature of the spontaneous fields emerging within the charge ordered state. 

Here we report muon-spin relaxation/rotation measurements of CsV$_3$Sb$_5$ in zero-, longitudinal, and transverse magnetic fields. We observe an  increase in the $\mu$SR relaxation rate upon cooling in zero-field, which is significantly enhanced below around 50~K, well below $T^* \sim 93$~K. This enhancement persists in a 50~G longitudinal field, pointing to the dynamic nature of these spontaneous fields. Measurements in the superconducting state indicate the lack of additional spontaneous fields appearing below the superconducting transition, and the superfluid density derived from transverse field results is consistent with the previously observed multiband superconductivity.

\section{Experimental details}
Polycrystalline samples of CsV$_3$Sb$_5$ were prepared by a solid-state reaction method. Stoichiometric amounts of Cs (liquid, Alfa 99.98\%), V (powder, Sigma 99.9\%), and Sb (shot, Alfa 99.999\%) were mixed thoroughly in a glove box filled with Ar gas. The mixture was subsequently loaded in an alumina crucible that was then jacketed in a tantalum tube. The tantalum tube was thereafter sealed in an evacuated quartz ampoule and heated up to 600$^\circ$C, held at this temperature for three days, before being furnace cooled to room temperature. The single phase of the as-grown samples was checked by powder x-ray diffraction on a PANalytical X-ray diffractometer with monochromatic Cu-K$\alpha_1$ radiation. $\mu$SR measurements were performed on powdered CsV$_3$Sb$_5$  samples at the ISIS facility at elevated temperatures between 5 and 180~K using the EMU spectrometer in a $^4$He-cryostat, and at low temperatures down to 0.1~K in a dilution refrigerator using the MuSR spectrometer. For zero-field (ZF) and longitudinal-field (LF) measurements, the asymmetry $A(t)$ between the number of positrons detected at the forward ($N_{\rm F}$) and backward  ($N_{\rm B}$)  positions was analyzed via 

\begin{equation}
A(t)=\frac{N_{\rm F}-\alpha N_{\rm B}}{N_{\rm F}+\alpha N_{\rm B}},
\label{Eq1}
\end{equation}

\noindent where $\alpha$ is a calibration constant. For transverse-field (TF) measurements, the time-dependent histograms corresponding to 16 groups of detectors were simultaneously analyzed using the \textsc{musrfit} software package \cite{Suter2012}.

\section{Results and Discussion}

\subsection{Zero- and longitudinal- field $\mu$SR at elevated temperatures}

\begin{figure}[tb]
\begin{center}
  \includegraphics[width=\columnwidth]{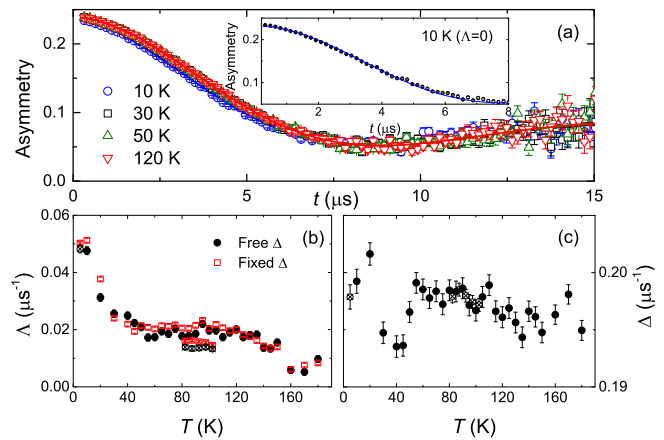}
  \end{center}
  \caption{(Color online) (a) Zero-field $\mu$SR measurements of CsV$_3$Sb$_5$ measured at four temperatures, where the solid lines show the results from fitting using Eq.~\ref{eqn:KT_ZFequation}. The inset shows the data at 10~K, where the solid line shows the results of fitting using Eq.~\ref{eqn:KT_ZFequation} with $\Lambda=0$. It can be seen that in the inset there is considerable deviation of the fitting from the data, showing the necessity of considering an exponential relaxation channel. The temperature dependence of the fitted values of (b) the exponential relaxation rate $\Lambda$, and (c) the Kubo-Toyabe relaxation rate $\Delta$. The values of $\Lambda$ are shown for the cases both when $\Delta$ is  fitted as a free parameter, as well as when $\Delta$ was fixed to the value obtained at 120~K. The symbols filled with a cross correspond to points which were measured in ZF after a LF had been applied, as described in the text.}
 \label{fig1}
\end{figure}

\begin{figure*}[htb]
\begin{center}
  \includegraphics[width=0.98\textwidth]{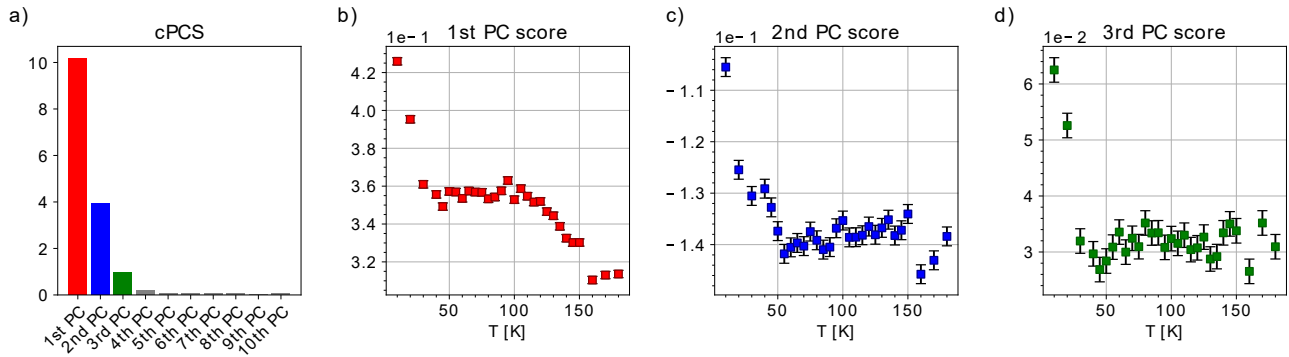}
  \end{center}
  \caption{(Color online) PCA of the ZF-$\mu$SR spectra of CsV$_3$Sb$_5$ in the normal state. (a) Cumulative principal component score for each PC. The first three PCs have the largest contribution to data reconstruction. (b)-- (d) Principal component scores for the 1st, 2nd and 3rd most important PCs as functions of temperature.}
 \label{fig:pcafig}
\end{figure*}

Figure~\ref{fig1}(a) displays the $\mu$SR spectra at several selected temperatures, measured in zero applied field on the EMU spectrometer. It can be seen that upon reducing the temperature, there is an increase in the relaxation rate of the asymmetry, indicating a broadening of the internal field distribution at the muon-stopping site. The shape of the muon spectra is indicative of a Kubo-Toyabe relaxation function, arising from a Gaussian distribution of magnetic fields static on the timescale of the muon lifetime (2.2$\mu s$). The data were analyzed taking into account both Kubo-Toyabe and exponential relaxation channels via

\begin{equation}\label{eqn:KT_ZFequation}
A(t)=A\left[\frac{1}{3}+\frac{2}{3}(1-\Delta^2t^2)e^{-\Delta^2t^2/2}\right]e^{-\Lambda t} + A_{\rm BG},
\end{equation}

\noindent where $A$ and $A_{\rm BG}$ correspond to the initial asymmetries for muons stopping in the sample and silver sample holder, respectively, which were fixed to the fitted values at 120~K, while $\Delta$ and $\Lambda$ are the Gaussian-Kubo-Toyabe and exponential  relaxation rates. Note that  if a purely Gaussian relaxation is considered ($\Lambda=0$), the low temperature spectra cannot be well fitted, as can be seen in the inset of Fig.~\ref{fig1}(a) where there is a clear deviation of such a fit from the data at 10~K. These results therefore point to the presence of both relaxation channels in CsV$_3$Sb$_5$. 

The temperature dependence of the fitted values of $\Delta$ and $\Lambda$ between 5 and 180~K are displayed in Figs.~\ref{fig1}(b) and (c) respectively. It can be seen that there is little change of $\Delta$ with temperature. On the other hand, at the highest temperatures $\Lambda$ is small, indicating that the relaxation is nearly entirely from the nuclear moments which are static on the timescale of the muon lifetime. Upon lowering the temperature there is an onset of the exponential component below around 160~K, with little change across the intermediate temperature range. Below around 50~K, there is a significant increase in $\Lambda$, which continues to increase with decreasing temperature down to the lowest measured temperature (5~K). To check that this low temperature increase of $\Lambda$ is not an artifact of correlations between the $\Lambda$ and $\Delta$ parameters, the data were fitted with $\Delta$ fixed to the value at 120~K ($\Delta=0.1961~\mu {\rm s}^{-1}$). As shown in Fig.~\ref{fig1}(b), a very similar trend is still observed in $\Lambda$, showing the intrinsic nature of this low temperature enhancement of the relaxation. Note that in Fig.~\ref{fig1}(b) a few points (centered with a cross) are systematically lower than the adjacent values. These were measured after a longitudinal field had been applied than removed at low temperatures, pointing to a weak dependence of this component on the field history. These results therefore show that there is a significant increase in the low temperature relaxation, which occurs well below the charge ordering transition at $T^*=93$~K. This is different to that observed in $\mu$SR measurements of  isostructural KV$_3$Sb$_5$  \cite{Mielke2021TRS} where an enhanced relaxation in the exponential channel onsets at $T^*=80$~K, but is similar to another study of CsV$_3$Sb$_5$ where the increase onsets at around 70~K, also below the charge ordering temperature, which is reported as evidence for a hidden flux phase \cite{yu2021evidence}. However, in Ref.~\cite{yu2021evidence} the relaxation corresponding to muons stopping in  CsV$_3$Sb$_5$ is ascribed as purely Gaussian, whereas here we find that both exponential and Gaussian components are required to account for our data, and the low temperature increase is predominantly in the exponential relaxation rate.

Since the analysis of the ZF-$\mu$SR data in the normal state of CsV$_3$Sb$_5$ using Eq.~\ref{eqn:KT_ZFequation} implies that both the Gaussian and Lorenzian relaxation channels are present, we complement this analysis by a recently introduced unbiased Principle Component analysis (PCA) technique \cite{Geron2019} based on unsupervised machine learning. In the PCA technique, a linear transformation in the data space is used to find only a few orthonormal basis vectors called Principal Components (PCs) such that they can capture the covariance of the data with respect to the average well. The projections of the original data on the PCs are called the principal component scores. This technique has recently been successfully used to identify transitions that are associated with TRS breaking in superconductors from ZF-$\mu$SR data of LaNiGa$_2$, and LaNi$_{1-x}$Cu$_x$C$_2$ \cite{Tula2021a} although the changes in these cases can be subtle, and magnetic transitions in antiferromagnetic BaFe$_2$Se$_2$O. We note that although this method can successfully determine changes in the asymmetry function by corresponding changes in the principal component scores, it is difficult to associate those changes to any particular type of relaxation channel. 

We show the results of PCA in \fig{fig:pcafig} for CsV$_3$Sb$_5$ performed jointly on ZF-$\mu$SR data from several materials showing TRS breaking \cite{Tula2021a}. We first identify the PCs that have largest contribution to the covariance of CsV$_3$Sb$_5$ data by computing the Cumulative Principal Component Score (cPCS) metric \cite{Tula2021b}. The cumulative PC scores for CsV$_3$Sb$_5$ are shown in \fig{fig:pcafig}(a). We note that only the first three PCs are important and the first PC captures the most covariance of the data. We have computed the error bars in the PC scores by assuming that the errors come solely from the experimental errors of the asymmetry functions. The PC scores at a given temperature are defined as
\begin{equation}
    PC_{\textrm{score}}^{[n]}(T_i) = \sum_{j}^{M} PC^{[n]}(t_j) \times A(T_i, t_j),
\end{equation}
where $PC^{[n]}(t_j)$ is a value of $n$-th PC at time $t_j$. Assuming that errors coming from different time windows are not correlated, we obtain the PC score standard deviation to be
\begin{equation}
    \mbox{SD}\left[PC_{\textrm{score}}^{[n]}(T_i) \right] = \sqrt{\sum_{j}^{M} \left(PC^{[n]}(t_j) \times E(T_i, t_j)\right)^2}
\end{equation}
where $E(T_i, t_j)$ indicates experimental error of asymmetry function $A(T_i, t_j)$ and we also assume that error coming from the variation of $PC^{[n]}(t_j)$ is negligible.

We note from the variations of the PC scores as a function of temperature shown in \fig{fig:pcafig}(b)--(d) that all the three PCs show a clear and significant change below $T \approx 50$~K. The changes in the PC scores set in below $T \approx 160$~K where in the intermediate temperature range there is little variation in the second and third PCs, while the first PC has a clear shoulder at around 100~K close to $T^*$, suggesting a relationship between the additional fields and the charge ordering. Thus the PCA corroborates the  results of the previous analysis of ZF-$\mu$SR data using Eq.~\ref{eqn:KT_ZFequation} showing the onset of addition fields below $T \approx 50$~K, which may be associated with spontaneous fields in the charge ordered state.

\begin{figure}[t]
\begin{center}
  \includegraphics[width=0.95\columnwidth]{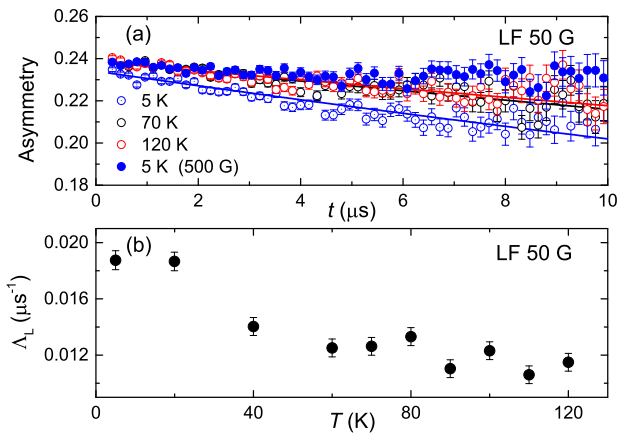}
  \end{center}
  \caption{(Color online) (a) $\mu$SR measurements of CsV$_3$Sb$_5$ performed in longitudinal applied fields, where the open symbols correspond to data at three temperatures in a longitudinal field of 50~G, while the filled symbols correspond to measurements in a 500~G field, where the muon is decoupled from the local fields. The solid lines show the results from fitting the 50~G data using Eq.~\ref{Eq3}. (b) Temperature dependence of the exponential relaxation rate $\Lambda_{\rm L}$ obtained from analyzing the data in the 50~G LF.}
 \label{fig:lfmusr}
\end{figure}

While an exponential relaxation is often interpreted as arising from fluctuating internal fields, it can also correspond to a Lorentizan distribution of static fields, as reported for KV$_3$Sb$_5$ \cite{Mielke2021TRS}, as well as several time-reversal symmetry breaking superconductors \cite{Luke1998,Hillier2009,Grinenko2020,Biswas2013,Shang2020}. To distinguish between these two scenarios, we performed measurements in a longitudinal field of 50~G, which are displayed for three temperatures in Fig.~\ref{fig:lfmusr}(a). It can be seen that a significant relaxing component is still observed in this longitudinal field, and the relaxation rate increases with decreasing temperature, while the Kubo-Toyabe component observed in ZF is absent, as expected when the muon is decoupled from the local field. When a larger LF of 500~G is applied, the muons are nearly entirely decoupled from the local field even at 5~K, with only a very small drop in the asymmetry. The 50G data were analyzed using 

\begin{equation}\label{Eq3}
A(t)=Ae^{-\Lambda_{\rm L} t} + A_{\rm BG},
\end{equation}

\noindent where $A_{\rm BG}$ was fixed to the same value as the ZF analysis. The temperature dependence of $\Lambda_{\rm L} $ in the 50~G LF is displayed in Fig.~\ref{fig:lfmusr}(b), where a sizeable value is still observed at low and intermediate temperatures. Furthermore, the significant low temperature enhancement observed below 50~K in the ZF data, is still present in a LF of 50~G. If the relatively small internal fields detected in the ZF measurements were purely static, they would be expected to be entirely decoupled by the LF, as was observed for KV$_3$Sb$_5$ in a 25~G LF \cite{kenney2020absence}, and this therefore suggests that the exponential  component for CsV$_3$Sb$_5$ corresponds to fluctuating magnetic fields, and as such, the low temperature enhancement also has a dynamic nature. The  dependence of $\Lambda_{\rm L} $ on the applied LF ($B_{\rm L}$) in the fast fluctuation limit can be estimated using the Redfield formula

\begin{equation}\label{Eq4}
\Lambda_{\rm L} = \frac{\Lambda\nu^2}{\gamma_{\mu}^2B_{\rm L}^2+\nu^2}
\end{equation}

\noindent where $\Lambda$ is the value in zero-field. Using the 5~K values of $\Lambda=0.04825~\mu {\rm s}^{-1}$, and $\Lambda_{\rm L}=0.01875~\mu {\rm s}^{-1}$ for the 50~G LF, we estimate $\nu=3.39~\mu {\rm s}^{-1}$, corresponding to a correlation time $\tau_c=0.29~\mu$s. The magnitude of the local fields $B_{\rm loc}$ can be related to $\Lambda$ via  $\Lambda=2(\gamma_{\mu}B_{\rm loc})^2/\nu$, yielding $B_{\rm loc}\approx 3$~G. These are similar local fields and correlation times inferred for the dynamic part of the $\mu$SR spectra of PrOs$_4$Sb$_{12}$ \cite{Aoki2003}, which were suggested to reflect the $4f$-electron dynamics of the system.

Due to the intriguing dynamic nature of the spontaneous fields below $~50$~K inside the CDW phase, it is difficult to associate them with a hidden static order, such as the one proposed in Ref.~\cite{yu2021evidence}. This further validates the unconventional nature of the CDW phase in CsV$_3$Sb$_5$ reported by other measurements such as pump-probe spectroscopy \cite{Ratcliff2021,Wang2021}. It was found that there is a change in rotational symmetry below 50~K from C$_6$ to C$_2$ which is more like a crossover and, the electronic and orbital degrees of freedom curiously behave differently in this regime \cite{Wang2021}. In addition, our measurements do not show the second  transition below $~30$~K reported in  Ref.~\cite{yu2021evidence}. Further experimental and theoretical studies are therefore necessary to uncover the true nature of the spontaneous fields associated with the CDW phase in CsV$_3$Sb$_5$. We note that in 1T-TaS$_2$, changes in the LF-$\mu$SR spectra at difference temperatures within the CDW phase were associated with the diffusion of spinons in distinct quantum spin liquid states \cite{manas2021quantum}. In addition, the dynamic fields observed in CsV$_3$Sb$_5$ are distinct from the static spontaneous fields inferred to appear at the charge order transition of isostructural KV$_3$Sb$_5$ \cite{kenney2020absence,Mielke2021TRS}. This points to fundamental differences in the charge ordered states of the two compounds, which could potentially arise from the presence of different types of van Hove singularity \cite{lin2021kagome}.

\subsection{$\mu$SR measurements in the superconducting state}

\begin{figure}[t]
\begin{center}
  \includegraphics[width=0.8\columnwidth]{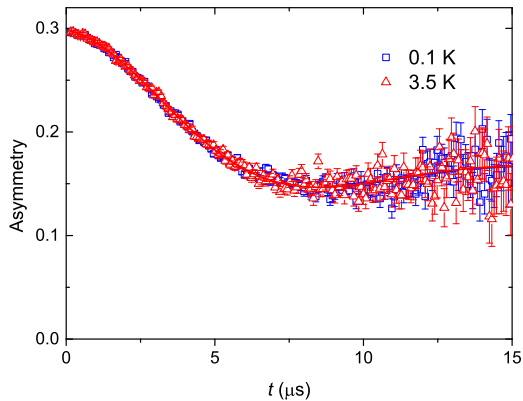}
  \end{center}
  \caption{(Color online) Zero-field $\mu$SR measurements performed on the MuSR spectrometer both in the normal state at 3.5~K, and far below the superconducting transition at 0.1~K. The solid lines show the results from fitting using Eq.~\ref{eqn:KT_ZFequation}. }
 \label{fig:zfmusr}
\end{figure}

In order to search for the appearance of spontaneous magnetic fields in the superconducting state, indicative of a TRS breaking pairing state \cite{Ghosh2020}, ZF $\mu$SR was measured using the MuSR spectrometer with the sample cooled in a dilution refrigerator. Figure~\ref{fig:zfmusr} displays spectra measured in ZF at two temperatures, at 3.5 ~K in the normal state, and at 0.1~K well below $T_{\rm c}$, where there is little difference between the measurements at the two temperatures. The data were fitted using Eq.~\ref{eqn:KT_ZFequation}, yielding $\Delta=0.205(2)~\mu {\rm s}^{-1}$ and $\Lambda=0.039(3)~\mu {\rm s}^{-1}$ for both temperatures. These suggest a lack of additional magnetic fields in the superconducting state, indicating that the pairing state does not break TRS, consistent with previous results for both CsV$_3$Sb$_5$ and KV$_3$Sb$_5$  \cite{gupta2021microscopic,Mielke2021TRS}. 

\begin{figure}[t]
\begin{center}
  \includegraphics[width=\columnwidth]{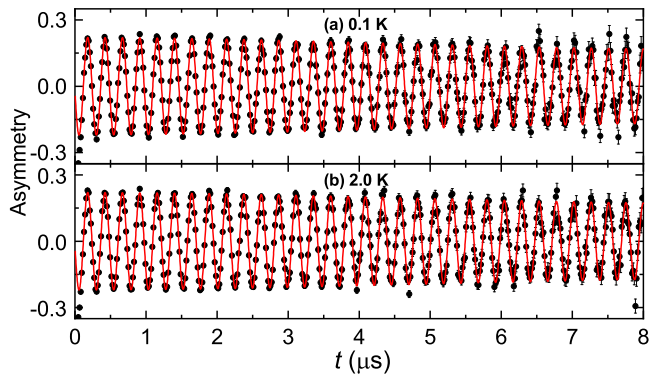}
  \end{center}
  \caption{(Color online) Muon-spin rotation measurements of CsV$_3$Sb measured in a 300~G transverse field at (a) 0.1~K, and (b) 2~K. The solid lines show the results from fitting using Eq.~\ref{equation5}.}
 \label{fig:tfmusr}
\end{figure}

\begin{figure}[t]
\begin{center}
  \includegraphics[width=0.8\columnwidth]{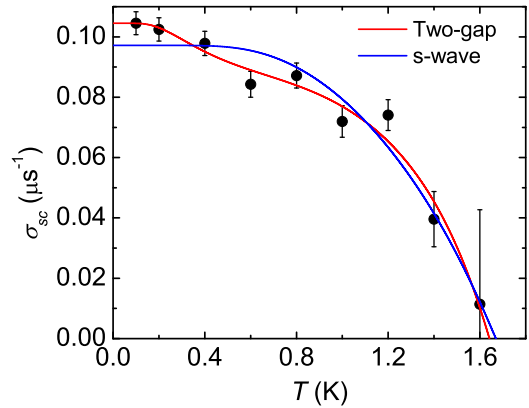}
  \end{center}
  \caption{(Color online) Temperature dependence of $\sigma_{\rm sc}$ of CsV$_3$Sb$_5$ determined from TF-$\mu$SR results, which is proportional to the superfluid density. The solid lines show the results from fitting with a two-gap $s$-wave model and an isotropic s-wave model as described in the text.}
 \label{fig:sigmasc}
\end{figure}

To determine the temperature dependence of the superfluid density, muon-spin rotation measurements were performed in a transverse field of 300~G, as shown for two temperatures in Fig.~\ref{fig:tfmusr}. An increased relaxation in the superconducting state corresponds to the formation of a flux-line lattice, which is sensitive to the magnitude of the penetration depth, and hence the superfluid density. The data were analyzed using 

\begin{equation}\label{equation5}
A_{\rm TF}(t)=A_se^{-\sigma^2t^2/2}\cos(\gamma_{\mu}B_st+\phi)+A_{bg}\cos(\gamma_{\mu}B_{bg}t+\phi)
\end{equation}

\noindent where the first and second terms correspond to muons stopping the sample and sample holder, respectively, and $\phi$ is a common phase. Here the ratio $A_s/A_{bg}$ was fixed to the fitted value obtained from fitting at the lowest measured temperature. The superconducting contribution to the relaxation $\sigma_{\rm sc}$ was calculated using $\sigma_{\rm sc}=\sqrt{\sigma^2-\sigma_{\rm n}^2}$, using a normal state contribution $\sigma_{\rm n}=0.210~\mu {\rm s}^{-1}$ estimated from the normal state data. The temperature dependence of $\sigma_{\rm sc}$, which is proportional to the superfluid density, is displayed in Fig.~\ref{fig:sigmasc}. A clear low temperature saturation of $\sigma_{\rm sc}$ cannot be resolved, and the data cannot be accounted for using a single gap $s$-wave model. On the other hand, there are not sufficient low temperature datapoints to resolve whether at low temperatures the data do approach a constant value, as expected for a fully open superconducting gap, or whether it exhibits power law behavior characteristic of nodal superconductivity. In a previous penetration depth study using the tunnel-diode oscillator based method \cite{duan2021nodeless}, fully gapped behavior is revealed by $\lambda(T)$ becoming flat only  at very low temperatures, below around 0.2~K, and the data are analyzed using an isotropic two-gap $s$-wave model. Similarly, we fitted $\sigma_{\rm sc}$  with the same two gap $s+s$ model, which as shown by the solid line in  Fig.~\ref{fig:sigmasc}, can well describe the data, with zero temperature gap magnitudes of $\Delta_1=0.55k_{\rm B}T_{\rm c}$ and $\Delta_2=2.78k_{\rm B}T_{\rm c}$, with a fraction corresponding to the smaller gap of $22\%$. Here the value of the small gap is very close to that from the analysis of the TDO data  \cite{duan2021nodeless}, while $\Delta_2$ is larger, being closer to that deduced previously from $\mu$SR \cite{gupta2021microscopic}. Therefore these results are consistent with the previous findings of two-gap superconductivity in CsV$_3$Sb$_5$ \cite{duan2021nodeless,gupta2021microscopic}.

The multigap superconductivity with TRS preserved found in CsV$_3$Sb$_5$ constrains the forms of possible superconducting instabilities in conjunction with other measurements which indicate isotropic full gap behavior  \cite{duan2021nodeless,song2021orbital,gupta2021microscopic,Yin2021,Liang2021,Xu2021,Mu2021}. In particular, the proposed nodal $f$-wave \cite{wu2021nature} type superconducting order parameter seems to be incompatible with the experimental observations in CsV$_3$Sb$_5$. Rather, a fully-gapped superconducting state resulting from the multiband nature and an intricate interplay \cite{lin2021kagome} between the CDW phase, superconductivity and topological order is expected to be realized in CsV$_3$Sb$_5$, and further studies are necessary to uncover its true nature and pairing mechanism.

\section{Conclusion}

In summary, we performed ZF, LF, and TF $\mu$SR measurements on the kagome lattice superconductor CsV$_3$Sb$_5$. Upon lowering the temperature, a significant increase in the relaxation rate corresponding to the exponential relaxation channel of  the ZF asymmetry is observed, which onsets below around 50~K, well below the charge ordering temperature $T^*=93$~K. Upon measuring in a LF of 50~G, a sizable relaxation is still observed, which also shows a similar low temperature increase, indicating the dynamic nature of these small fields. Meanwhile ZF measurements at lower temperatures show no detectable change upon entering the superconducting state, indicating that the superconducting order parameter does not break TRS, while the TF $\mu$SR analysis is consistent with the previous findings of two-gap $s$-wave superconductivity. While it is still an open question as to whether the additional internal fields onsetting well below $T^*$ are related to changes in the charge ordered state, the dynamic nature of these fields inferred from our study suggests that they cannot be straightforwardly interpreted in terms of static spontaneous fields arising from the unusual charge ordered state.

\section{acknowledgments}

This work was supported by the National Key R$\&$D Program of China (2017YFA0303100), the Key R$\&$D Program of Zhejiang Province, China (2021C01002), the National Natural Science Foundation of China (11874320, 12034017, 11974306 and 11974061), and the Zhejiang Provincial Natural Science Foundation of China (R22A0410240). SKG acknowledges the Leverhulme Trust for support through the Leverhulme early career fellowship and thanks J. Quintanilla for discussions. Experiments at the ISIS Pulsed Neutron and Muon Source were supported by a beamtime allocation from the Science and Technology Facilities Council (RB2000246 \cite{RB1} and RB2000245  \cite{RB2})


\begin{thebibliography}{56}%
\makeatletter
\providecommand \@ifxundefined [1]{%
 \@ifx{#1\undefined}
}%
\providecommand \@ifnum [1]{%
 \ifnum #1\expandafter \@firstoftwo
 \else \expandafter \@secondoftwo
 \fi
}%
\providecommand \@ifx [1]{%
 \ifx #1\expandafter \@firstoftwo
 \else \expandafter \@secondoftwo
 \fi
}%
\providecommand \natexlab [1]{#1}%
\providecommand \enquote  [1]{``#1''}%
\providecommand \bibnamefont  [1]{#1}%
\providecommand \bibfnamefont [1]{#1}%
\providecommand \citenamefont [1]{#1}%
\providecommand \href@noop [0]{\@secondoftwo}%
\providecommand \href [0]{\begingroup \@sanitize@url \@href}%
\providecommand \@href[1]{\@@startlink{#1}\@@href}%
\providecommand \@@href[1]{\endgroup#1\@@endlink}%
\providecommand \@sanitize@url [0]{\catcode `\\12\catcode `\$12\catcode
  `\&12\catcode `\#12\catcode `\^12\catcode `\_12\catcode `\%12\relax}%
\providecommand \@@startlink[1]{}%
\providecommand \@@endlink[0]{}%
\providecommand \url  [0]{\begingroup\@sanitize@url \@url }%
\providecommand \@url [1]{\endgroup\@href {#1}{\urlprefix }}%
\providecommand \urlprefix  [0]{URL }%
\providecommand \Eprint [0]{\href }%
\providecommand \doibase [0]{https://doi.org/}%
\providecommand \selectlanguage [0]{\@gobble}%
\providecommand \bibinfo  [0]{\@secondoftwo}%
\providecommand \bibfield  [0]{\@secondoftwo}%
\providecommand \translation [1]{[#1]}%
\providecommand \BibitemOpen [0]{}%
\providecommand \bibitemStop [0]{}%
\providecommand \bibitemNoStop [0]{.\EOS\space}%
\providecommand \EOS [0]{\spacefactor3000\relax}%
\providecommand \BibitemShut  [1]{\csname bibitem#1\endcsname}%
\let\auto@bib@innerbib\@empty
\bibitem [{\citenamefont {Balents}(2010)}]{Balents2010}%
  \BibitemOpen
  \bibfield  {author} {\bibinfo {author} {\bibfnamefont {L.}~\bibnamefont
  {Balents}},\ }\bibfield  {title} {\bibinfo {title} {Spin liquids in
  frustrated magnets},\ }\href {https://doi.org/10.1038/nature08917} {\bibfield
   {journal} {\bibinfo  {journal} {Nature}\ }\textbf {\bibinfo {volume}
  {464}},\ \bibinfo {pages} {199} (\bibinfo {year} {2010})}\BibitemShut
  {NoStop}%
\bibitem [{\citenamefont {Fu}\ \emph {et~al.}(2015)\citenamefont {Fu},
  \citenamefont {Imai}, \citenamefont {Han},\ and\ \citenamefont
  {Lee}}]{Mingxuan2015}%
  \BibitemOpen
  \bibfield  {author} {\bibinfo {author} {\bibfnamefont {M.}~\bibnamefont
  {Fu}}, \bibinfo {author} {\bibfnamefont {T.}~\bibnamefont {Imai}}, \bibinfo
  {author} {\bibfnamefont {T.-H.}\ \bibnamefont {Han}}, and\ \bibinfo {author}
  {\bibfnamefont {Y.~S.}\ \bibnamefont {Lee}},\ }\bibfield  {title} {\bibinfo
  {title} {Evidence for a gapped spin-liquid ground state in a kagome
  {H}eisenberg antiferromagnet},\ }\href
  {https://doi.org/10.1126/science.aab2120} {\bibfield  {journal} {\bibinfo
  {journal} {Science}\ }\textbf {\bibinfo {volume} {350}},\ \bibinfo {pages}
  {655} (\bibinfo {year} {2015})}\BibitemShut {NoStop}%
\bibitem [{\citenamefont {Han}\ \emph {et~al.}(2012)\citenamefont {Han},
  \citenamefont {Helton}, \citenamefont {Chu}, \citenamefont {Nocera},
  \citenamefont {Rodriguez-Rivera}, \citenamefont {Broholm},\ and\
  \citenamefont {Lee}}]{Han2012}%
  \BibitemOpen
  \bibfield  {author} {\bibinfo {author} {\bibfnamefont {T.-H.}\ \bibnamefont
  {Han}}, \bibinfo {author} {\bibfnamefont {J.~S.}\ \bibnamefont {Helton}},
  \bibinfo {author} {\bibfnamefont {S.}~\bibnamefont {Chu}}, \bibinfo {author}
  {\bibfnamefont {D.~G.}\ \bibnamefont {Nocera}}, \bibinfo {author}
  {\bibfnamefont {J.~A.}\ \bibnamefont {Rodriguez-Rivera}}, \bibinfo {author}
  {\bibfnamefont {C.}~\bibnamefont {Broholm}}, and\ \bibinfo {author}
  {\bibfnamefont {Y.~S.}\ \bibnamefont {Lee}},\ }\bibfield  {title} {\bibinfo
  {title} {Fractionalized excitations in the spin-liquid state of a
  kagome-lattice antiferromagnet},\ }\href
  {https://doi.org/10.1038/nature11659} {\bibfield  {journal} {\bibinfo
  {journal} {Nature}\ }\textbf {\bibinfo {volume} {492}},\ \bibinfo {pages}
  {406} (\bibinfo {year} {2012})}\BibitemShut {NoStop}%
\bibitem [{\citenamefont {Lin}\ \emph {et~al.}(2018)\citenamefont {Lin},
  \citenamefont {Choi}, \citenamefont {Zhang}, \citenamefont {Qin},
  \citenamefont {Yi}, \citenamefont {Wang}, \citenamefont {Li}, \citenamefont
  {Wang}, \citenamefont {Zhang}, \citenamefont {Sun}, \citenamefont {Wei},
  \citenamefont {Zhang}, \citenamefont {Guo}, \citenamefont {Lu}, \citenamefont
  {Cho}, \citenamefont {Zeng},\ and\ \citenamefont {Zhang}}]{Lin2018}%
  \BibitemOpen
  \bibfield  {author} {\bibinfo {author} {\bibfnamefont {Z.}~\bibnamefont
  {Lin}}, \bibinfo {author} {\bibfnamefont {J.-H.}\ \bibnamefont {Choi}},
  \bibinfo {author} {\bibfnamefont {Q.}~\bibnamefont {Zhang}}, \bibinfo
  {author} {\bibfnamefont {W.}~\bibnamefont {Qin}}, \bibinfo {author}
  {\bibfnamefont {S.}~\bibnamefont {Yi}}, \bibinfo {author} {\bibfnamefont
  {P.}~\bibnamefont {Wang}}, \bibinfo {author} {\bibfnamefont {L.}~\bibnamefont
  {Li}}, \bibinfo {author} {\bibfnamefont {Y.}~\bibnamefont {Wang}}, \bibinfo
  {author} {\bibfnamefont {H.}~\bibnamefont {Zhang}}, \bibinfo {author}
  {\bibfnamefont {Z.}~\bibnamefont {Sun}}, \bibinfo {author} {\bibfnamefont
  {L.}~\bibnamefont {Wei}}, \bibinfo {author} {\bibfnamefont {S.}~\bibnamefont
  {Zhang}}, \bibinfo {author} {\bibfnamefont {T.}~\bibnamefont {Guo}}, \bibinfo
  {author} {\bibfnamefont {Q.}~\bibnamefont {Lu}}, \bibinfo {author}
  {\bibfnamefont {J.-H.}\ \bibnamefont {Cho}}, \bibinfo {author} {\bibfnamefont
  {C.}~\bibnamefont {Zeng}}, and\ \bibinfo {author} {\bibfnamefont
  {Z.}~\bibnamefont {Zhang}},\ }\bibfield  {title} {\bibinfo {title} {Flatbands
  and emergent ferromagnetic ordering in {${\mathrm{Fe}}_{3}{\mathrm{Sn}}_{2}$}
  kagome lattices},\ }\href {https://doi.org/10.1103/PhysRevLett.121.096401}
  {\bibfield  {journal} {\bibinfo  {journal} {Phys. Rev. Lett.}\ }\textbf
  {\bibinfo {volume} {121}},\ \bibinfo {pages} {096401} (\bibinfo {year}
  {2018})}\BibitemShut {NoStop}%
\bibitem [{\citenamefont {Yin}\ \emph {et~al.}(2019)\citenamefont {Yin},
  \citenamefont {Zhang}, \citenamefont {Chang}, \citenamefont {Wang},
  \citenamefont {Tsirkin}, \citenamefont {Guguchia}, \citenamefont {Lian},
  \citenamefont {Zhou}, \citenamefont {Jiang}, \citenamefont {Belopolski} \emph
  {et~al.}}]{Yin2019}%
  \BibitemOpen
  \bibfield  {author} {\bibinfo {author} {\bibfnamefont {J.-X.}\ \bibnamefont
  {Yin}}, \bibinfo {author} {\bibfnamefont {S.~S.}\ \bibnamefont {Zhang}},
  \bibinfo {author} {\bibfnamefont {G.}~\bibnamefont {Chang}}, \bibinfo
  {author} {\bibfnamefont {Q.}~\bibnamefont {Wang}}, \bibinfo {author}
  {\bibfnamefont {S.~S.}\ \bibnamefont {Tsirkin}}, \bibinfo {author}
  {\bibfnamefont {Z.}~\bibnamefont {Guguchia}}, \bibinfo {author}
  {\bibfnamefont {B.}~\bibnamefont {Lian}}, \bibinfo {author} {\bibfnamefont
  {H.}~\bibnamefont {Zhou}}, \bibinfo {author} {\bibfnamefont {K.}~\bibnamefont
  {Jiang}}, \bibinfo {author} {\bibfnamefont {I.}~\bibnamefont {Belopolski}},
  \emph {et~al.},\ }\bibfield  {title} {\bibinfo {title} {Negative flat band
  magnetism in a spin--orbit-coupled correlated kagome magnet},\ }\href
  {https://doi.org/10.1038/s41567-019-0426-7} {\bibfield  {journal} {\bibinfo
  {journal} {Nature Physics}\ }\textbf {\bibinfo {volume} {15}},\ \bibinfo
  {pages} {443} (\bibinfo {year} {2019})}\BibitemShut {NoStop}%
\bibitem [{\citenamefont {Kang}\ \emph {et~al.}(2019)\citenamefont {Kang},
  \citenamefont {Ye}, \citenamefont {Fang}, \citenamefont {You}, \citenamefont
  {Levitan}, \citenamefont {Han}, \citenamefont {Facio}, \citenamefont
  {Jozwiak}, \citenamefont {Bostwick}, \citenamefont {Rotenberg}, \citenamefont
  {Chan}, \citenamefont {McDonald}, \citenamefont {Graf}, \citenamefont
  {Kaznatcheev}, \citenamefont {Vescovo}, \citenamefont {Bell}, \citenamefont
  {Kaxiras}, \citenamefont {van~den Brink}, \citenamefont {Richter},
  \citenamefont {Ghimire}, \citenamefont {Checkelsky},\ and\ \citenamefont
  {Comin}}]{Kang2019}%
  \BibitemOpen
  \bibfield  {author} {\bibinfo {author} {\bibfnamefont {M.}~\bibnamefont
  {Kang}}, \bibinfo {author} {\bibfnamefont {L.}~\bibnamefont {Ye}}, \bibinfo
  {author} {\bibfnamefont {S.}~\bibnamefont {Fang}}, \bibinfo {author}
  {\bibfnamefont {J.-S.}\ \bibnamefont {You}}, \bibinfo {author} {\bibfnamefont
  {A.}~\bibnamefont {Levitan}}, \bibinfo {author} {\bibfnamefont
  {M.}~\bibnamefont {Han}}, \bibinfo {author} {\bibfnamefont {J.~I.}\
  \bibnamefont {Facio}}, \bibinfo {author} {\bibfnamefont {C.}~\bibnamefont
  {Jozwiak}}, \bibinfo {author} {\bibfnamefont {A.}~\bibnamefont {Bostwick}},
  \bibinfo {author} {\bibfnamefont {E.}~\bibnamefont {Rotenberg}}, \bibinfo
  {author} {\bibfnamefont {M.~K.}\ \bibnamefont {Chan}}, \bibinfo {author}
  {\bibfnamefont {R.~D.}\ \bibnamefont {McDonald}}, \bibinfo {author}
  {\bibfnamefont {D.}~\bibnamefont {Graf}}, \bibinfo {author} {\bibfnamefont
  {K.}~\bibnamefont {Kaznatcheev}}, \bibinfo {author} {\bibfnamefont
  {E.}~\bibnamefont {Vescovo}}, \bibinfo {author} {\bibfnamefont {D.~C.}\
  \bibnamefont {Bell}}, \bibinfo {author} {\bibfnamefont {E.}~\bibnamefont
  {Kaxiras}}, \bibinfo {author} {\bibfnamefont {J.}~\bibnamefont {van~den
  Brink}}, \bibinfo {author} {\bibfnamefont {M.}~\bibnamefont {Richter}},
  \bibinfo {author} {\bibfnamefont {M.~P.}\ \bibnamefont {Ghimire}}, \bibinfo
  {author} {\bibfnamefont {J.~G.}\ \bibnamefont {Checkelsky}}, and\ \bibinfo
  {author} {\bibfnamefont {R.}~\bibnamefont {Comin}},\ }\bibfield  {title}
  {\bibinfo {title} {Dirac fermions and flat bands in the ideal kagome metal
  {FeSn}},\ }\href {https://doi.org/10.1038/s41563-019-0531-0} {\bibfield
  {journal} {\bibinfo  {journal} {Nature Materials}\ }\textbf {\bibinfo
  {volume} {19}},\ \bibinfo {pages} {163} (\bibinfo {year} {2019})}\BibitemShut
  {NoStop}%
\bibitem [{\citenamefont {Guo}\ and\ \citenamefont {Franz}(2009)}]{Guo2009}%
  \BibitemOpen
  \bibfield  {author} {\bibinfo {author} {\bibfnamefont {H.-M.}\ \bibnamefont
  {Guo}} and\ \bibinfo {author} {\bibfnamefont {M.}~\bibnamefont {Franz}},\
  }\bibfield  {title} {\bibinfo {title} {Topological insulator on the kagome
  lattice},\ }\href {https://doi.org/10.1103/PhysRevB.80.113102} {\bibfield
  {journal} {\bibinfo  {journal} {Phys. Rev. B}\ }\textbf {\bibinfo {volume}
  {80}},\ \bibinfo {pages} {113102} (\bibinfo {year} {2009})}\BibitemShut
  {NoStop}%
\bibitem [{\citenamefont {Ortiz}\ \emph {et~al.}(2019)\citenamefont {Ortiz},
  \citenamefont {Gomes}, \citenamefont {Morey}, \citenamefont {Winiarski},
  \citenamefont {Bordelon}, \citenamefont {Mangum}, \citenamefont {Oswald},
  \citenamefont {Rodriguez-Rivera}, \citenamefont {Neilson}, \citenamefont
  {Wilson}, \citenamefont {Ertekin}, \citenamefont {McQueen},\ and\
  \citenamefont {Toberer}}]{Ortiz2019}%
  \BibitemOpen
  \bibfield  {author} {\bibinfo {author} {\bibfnamefont {B.~R.}\ \bibnamefont
  {Ortiz}}, \bibinfo {author} {\bibfnamefont {L.~C.}\ \bibnamefont {Gomes}},
  \bibinfo {author} {\bibfnamefont {J.~R.}\ \bibnamefont {Morey}}, \bibinfo
  {author} {\bibfnamefont {M.}~\bibnamefont {Winiarski}}, \bibinfo {author}
  {\bibfnamefont {M.}~\bibnamefont {Bordelon}}, \bibinfo {author}
  {\bibfnamefont {J.~S.}\ \bibnamefont {Mangum}}, \bibinfo {author}
  {\bibfnamefont {I.~W.~H.}\ \bibnamefont {Oswald}}, \bibinfo {author}
  {\bibfnamefont {J.~A.}\ \bibnamefont {Rodriguez-Rivera}}, \bibinfo {author}
  {\bibfnamefont {J.~R.}\ \bibnamefont {Neilson}}, \bibinfo {author}
  {\bibfnamefont {S.~D.}\ \bibnamefont {Wilson}}, \bibinfo {author}
  {\bibfnamefont {E.}~\bibnamefont {Ertekin}}, \bibinfo {author} {\bibfnamefont
  {T.~M.}\ \bibnamefont {McQueen}}, and\ \bibinfo {author} {\bibfnamefont
  {E.~S.}\ \bibnamefont {Toberer}},\ }\bibfield  {title} {\bibinfo {title} {New
  kagome prototype materials: discovery of
  {${\mathrm{KV}}_{3}{\mathrm{Sb}}_{5},{\mathrm{RbV}}_{3}{\mathrm{Sb}}_{5}$,
  and ${\mathrm{CsV}}_{3}{\mathrm{Sb}}_{5}$}},\ }\href
  {https://doi.org/10.1103/PhysRevMaterials.3.094407} {\bibfield  {journal}
  {\bibinfo  {journal} {Phys. Rev. Materials}\ }\textbf {\bibinfo {volume}
  {3}},\ \bibinfo {pages} {094407} (\bibinfo {year} {2019})}\BibitemShut
  {NoStop}%
\bibitem [{\citenamefont {Ortiz}\ \emph {et~al.}(2020)\citenamefont {Ortiz},
  \citenamefont {Teicher}, \citenamefont {Hu}, \citenamefont {Zuo},
  \citenamefont {Sarte}, \citenamefont {Schueller}, \citenamefont {Abeykoon},
  \citenamefont {Krogstad}, \citenamefont {Rosenkranz}, \citenamefont {Osborn},
  \citenamefont {Seshadri}, \citenamefont {Balents}, \citenamefont {He},\ and\
  \citenamefont {Wilson}}]{Ortiz2020}%
  \BibitemOpen
  \bibfield  {author} {\bibinfo {author} {\bibfnamefont {B.~R.}\ \bibnamefont
  {Ortiz}}, \bibinfo {author} {\bibfnamefont {S.~M.~L.}\ \bibnamefont
  {Teicher}}, \bibinfo {author} {\bibfnamefont {Y.}~\bibnamefont {Hu}},
  \bibinfo {author} {\bibfnamefont {J.~L.}\ \bibnamefont {Zuo}}, \bibinfo
  {author} {\bibfnamefont {P.~M.}\ \bibnamefont {Sarte}}, \bibinfo {author}
  {\bibfnamefont {E.~C.}\ \bibnamefont {Schueller}}, \bibinfo {author}
  {\bibfnamefont {A.~M.~M.}\ \bibnamefont {Abeykoon}}, \bibinfo {author}
  {\bibfnamefont {M.~J.}\ \bibnamefont {Krogstad}}, \bibinfo {author}
  {\bibfnamefont {S.}~\bibnamefont {Rosenkranz}}, \bibinfo {author}
  {\bibfnamefont {R.}~\bibnamefont {Osborn}}, \bibinfo {author} {\bibfnamefont
  {R.}~\bibnamefont {Seshadri}}, \bibinfo {author} {\bibfnamefont
  {L.}~\bibnamefont {Balents}}, \bibinfo {author} {\bibfnamefont
  {J.}~\bibnamefont {He}}, and\ \bibinfo {author} {\bibfnamefont {S.~D.}\
  \bibnamefont {Wilson}},\ }\bibfield  {title} {\bibinfo {title}
  {{$\mathrm{Cs}{\mathrm{V}}_{3}{\mathrm{Sb}}_{5}$: A ${\mathbb{Z}}_{2}$}
  topological kagome metal with a superconducting ground state},\ }\href
  {https://doi.org/10.1103/PhysRevLett.125.247002} {\bibfield  {journal}
  {\bibinfo  {journal} {Phys. Rev. Lett.}\ }\textbf {\bibinfo {volume} {125}},\
  \bibinfo {pages} {247002} (\bibinfo {year} {2020})}\BibitemShut {NoStop}%
\bibitem [{\citenamefont {Ortiz}\ \emph {et~al.}(2021)\citenamefont {Ortiz},
  \citenamefont {Sarte}, \citenamefont {Kenney}, \citenamefont {Graf},
  \citenamefont {Teicher}, \citenamefont {Seshadri},\ and\ \citenamefont
  {Wilson}}]{Ortiz2021}%
  \BibitemOpen
  \bibfield  {author} {\bibinfo {author} {\bibfnamefont {B.~R.}\ \bibnamefont
  {Ortiz}}, \bibinfo {author} {\bibfnamefont {P.~M.}\ \bibnamefont {Sarte}},
  \bibinfo {author} {\bibfnamefont {E.~M.}\ \bibnamefont {Kenney}}, \bibinfo
  {author} {\bibfnamefont {M.~J.}\ \bibnamefont {Graf}}, \bibinfo {author}
  {\bibfnamefont {S.~M.~L.}\ \bibnamefont {Teicher}}, \bibinfo {author}
  {\bibfnamefont {R.}~\bibnamefont {Seshadri}}, and\ \bibinfo {author}
  {\bibfnamefont {S.~D.}\ \bibnamefont {Wilson}},\ }\bibfield  {title}
  {\bibinfo {title} {Superconductivity in the {${\mathbb{Z}}_{2}$ kagome metal
  ${\mathrm{KV}}_{3}{\mathrm{Sb}}_{5}$}},\ }\href
  {https://doi.org/10.1103/PhysRevMaterials.5.034801} {\bibfield  {journal}
  {\bibinfo  {journal} {Phys. Rev. Materials}\ }\textbf {\bibinfo {volume}
  {5}},\ \bibinfo {pages} {034801} (\bibinfo {year} {2021})}\BibitemShut
  {NoStop}%
\bibitem [{\citenamefont {Yin}\ \emph {et~al.}(2021{\natexlab{a}})\citenamefont
  {Yin}, \citenamefont {Tu}, \citenamefont {Gong}, \citenamefont {Fu},
  \citenamefont {Yan},\ and\ \citenamefont {Lei}}]{yin2021superconductivity}%
  \BibitemOpen
  \bibfield  {author} {\bibinfo {author} {\bibfnamefont {Q.}~\bibnamefont
  {Yin}}, \bibinfo {author} {\bibfnamefont {Z.}~\bibnamefont {Tu}}, \bibinfo
  {author} {\bibfnamefont {C.}~\bibnamefont {Gong}}, \bibinfo {author}
  {\bibfnamefont {Y.}~\bibnamefont {Fu}}, \bibinfo {author} {\bibfnamefont
  {S.}~\bibnamefont {Yan}}, and\ \bibinfo {author} {\bibfnamefont
  {H.}~\bibnamefont {Lei}},\ }\bibfield  {title} {\bibinfo {title}
  {Superconductivity and normal-state properties of kagome metal
  {RbV$_3$Sb$_5$} single crystals},\ }\href
  {https://doi.org/10.1088/0256-307x/38/3/037403} {\bibfield  {journal}
  {\bibinfo  {journal} {Chinese Physics Letters}\ }\textbf {\bibinfo {volume}
  {38}},\ \bibinfo {pages} {037403} (\bibinfo {year}
  {2021}{\natexlab{a}})}\BibitemShut {NoStop}%
\bibitem [{\citenamefont {Yang}\ \emph {et~al.}(2020)\citenamefont {Yang},
  \citenamefont {Wang}, \citenamefont {Ortiz}, \citenamefont {Liu},
  \citenamefont {Gayles}, \citenamefont {Derunova}, \citenamefont
  {Gonzalez-Hernandez}, \citenamefont {{\v{S}}mejkal}, \citenamefont {Chen},
  \citenamefont {Parkin}, \citenamefont {Wilson}, \citenamefont {Toberer},
  \citenamefont {McQueen},\ and\ \citenamefont {Ali}}]{Yang2020}%
  \BibitemOpen
  \bibfield  {author} {\bibinfo {author} {\bibfnamefont {S.-Y.}\ \bibnamefont
  {Yang}}, \bibinfo {author} {\bibfnamefont {Y.}~\bibnamefont {Wang}}, \bibinfo
  {author} {\bibfnamefont {B.~R.}\ \bibnamefont {Ortiz}}, \bibinfo {author}
  {\bibfnamefont {D.}~\bibnamefont {Liu}}, \bibinfo {author} {\bibfnamefont
  {J.}~\bibnamefont {Gayles}}, \bibinfo {author} {\bibfnamefont
  {E.}~\bibnamefont {Derunova}}, \bibinfo {author} {\bibfnamefont
  {R.}~\bibnamefont {Gonzalez-Hernandez}}, \bibinfo {author} {\bibfnamefont
  {L.}~\bibnamefont {{\v{S}}mejkal}}, \bibinfo {author} {\bibfnamefont
  {Y.}~\bibnamefont {Chen}}, \bibinfo {author} {\bibfnamefont {S.~S.~P.}\
  \bibnamefont {Parkin}}, \bibinfo {author} {\bibfnamefont {S.~D.}\
  \bibnamefont {Wilson}}, \bibinfo {author} {\bibfnamefont {E.~S.}\
  \bibnamefont {Toberer}}, \bibinfo {author} {\bibfnamefont {T.}~\bibnamefont
  {McQueen}}, and\ \bibinfo {author} {\bibfnamefont {M.~N.}\ \bibnamefont
  {Ali}},\ }\bibfield  {title} {\bibinfo {title} {Giant, unconventional
  anomalous {H}all effect in the metallic frustrated magnet candidate,
  {KV$_{3}$Sb$_{5}$}},\ }\href {https://doi.org/10.1126/sciadv.abb6003}
  {\bibfield  {journal} {\bibinfo  {journal} {Science Advances}\ }\textbf
  {\bibinfo {volume} {6}},\ \bibinfo {pages} {eabb6003} (\bibinfo {year}
  {2020})}\BibitemShut {NoStop}%
\bibitem [{\citenamefont {Yu}\ \emph {et~al.}(2021{\natexlab{a}})\citenamefont
  {Yu}, \citenamefont {Wu}, \citenamefont {Wang}, \citenamefont {Lei},
  \citenamefont {Zhuo}, \citenamefont {Ying},\ and\ \citenamefont
  {Chen}}]{yu2021concurrence}%
  \BibitemOpen
  \bibfield  {author} {\bibinfo {author} {\bibfnamefont {F.~H.}\ \bibnamefont
  {Yu}}, \bibinfo {author} {\bibfnamefont {T.}~\bibnamefont {Wu}}, \bibinfo
  {author} {\bibfnamefont {Z.~Y.}\ \bibnamefont {Wang}}, \bibinfo {author}
  {\bibfnamefont {B.}~\bibnamefont {Lei}}, \bibinfo {author} {\bibfnamefont
  {W.~Z.}\ \bibnamefont {Zhuo}}, \bibinfo {author} {\bibfnamefont {J.~J.}\
  \bibnamefont {Ying}}, and\ \bibinfo {author} {\bibfnamefont {X.~H.}\
  \bibnamefont {Chen}},\ }\bibfield  {title} {\bibinfo {title} {Concurrence of
  anomalous {H}all effect and charge density wave in a superconducting
  topological kagome metal},\ }\href
  {https://doi.org/10.1103/PhysRevB.104.L041103} {\bibfield  {journal}
  {\bibinfo  {journal} {Phys. Rev. B}\ }\textbf {\bibinfo {volume} {104}},\
  \bibinfo {pages} {L041103} (\bibinfo {year}
  {2021}{\natexlab{a}})}\BibitemShut {NoStop}%
\bibitem [{\citenamefont {Jiang}\ \emph {et~al.}(2021)\citenamefont {Jiang},
  \citenamefont {Yin}, \citenamefont {Denner}, \citenamefont {Shumiya},
  \citenamefont {Ortiz}, \citenamefont {Xu}, \citenamefont {Guguchia},
  \citenamefont {He}, \citenamefont {Hossain}, \citenamefont {Liu},
  \citenamefont {Ruff}, \citenamefont {Kautzsch}, \citenamefont {Zhang},
  \citenamefont {Chang}, \citenamefont {Belopolski}, \citenamefont {Zhang},
  \citenamefont {Cochran}, \citenamefont {Multer}, \citenamefont {Litskevich},
  \citenamefont {Cheng}, \citenamefont {Yang}, \citenamefont {Wang},
  \citenamefont {Thomale}, \citenamefont {Neupert}, \citenamefont {Wilson},\
  and\ \citenamefont {Hasan}}]{jiang2020discovery}%
  \BibitemOpen
  \bibfield  {author} {\bibinfo {author} {\bibfnamefont {Y.-X.}\ \bibnamefont
  {Jiang}}, \bibinfo {author} {\bibfnamefont {J.-X.}\ \bibnamefont {Yin}},
  \bibinfo {author} {\bibfnamefont {M.~M.}\ \bibnamefont {Denner}}, \bibinfo
  {author} {\bibfnamefont {N.}~\bibnamefont {Shumiya}}, \bibinfo {author}
  {\bibfnamefont {B.~R.}\ \bibnamefont {Ortiz}}, \bibinfo {author}
  {\bibfnamefont {G.}~\bibnamefont {Xu}}, \bibinfo {author} {\bibfnamefont
  {Z.}~\bibnamefont {Guguchia}}, \bibinfo {author} {\bibfnamefont
  {J.}~\bibnamefont {He}}, \bibinfo {author} {\bibfnamefont {M.~S.}\
  \bibnamefont {Hossain}}, \bibinfo {author} {\bibfnamefont {X.}~\bibnamefont
  {Liu}}, \bibinfo {author} {\bibfnamefont {J.}~\bibnamefont {Ruff}}, \bibinfo
  {author} {\bibfnamefont {L.}~\bibnamefont {Kautzsch}}, \bibinfo {author}
  {\bibfnamefont {S.~S.}\ \bibnamefont {Zhang}}, \bibinfo {author}
  {\bibfnamefont {G.}~\bibnamefont {Chang}}, \bibinfo {author} {\bibfnamefont
  {I.}~\bibnamefont {Belopolski}}, \bibinfo {author} {\bibfnamefont
  {Q.}~\bibnamefont {Zhang}}, \bibinfo {author} {\bibfnamefont {T.~A.}\
  \bibnamefont {Cochran}}, \bibinfo {author} {\bibfnamefont {D.}~\bibnamefont
  {Multer}}, \bibinfo {author} {\bibfnamefont {M.}~\bibnamefont {Litskevich}},
  \bibinfo {author} {\bibfnamefont {Z.-J.}\ \bibnamefont {Cheng}}, \bibinfo
  {author} {\bibfnamefont {X.~P.}\ \bibnamefont {Yang}}, \bibinfo {author}
  {\bibfnamefont {Z.}~\bibnamefont {Wang}}, \bibinfo {author} {\bibfnamefont
  {R.}~\bibnamefont {Thomale}}, \bibinfo {author} {\bibfnamefont
  {T.}~\bibnamefont {Neupert}}, \bibinfo {author} {\bibfnamefont {S.~D.}\
  \bibnamefont {Wilson}}, and\ \bibinfo {author} {\bibfnamefont {M.~Z.}\
  \bibnamefont {Hasan}},\ }\bibfield  {title} {\bibinfo {title} {Unconventional
  chiral charge order in kagome superconductor {KV$_3$Sb$_5$}},\ }\href
  {https://doi.org/10.1038/s41563-021-01034-y} {\bibfield  {journal} {\bibinfo
  {journal} {Nature Materials}\ }\textbf {\bibinfo {volume} {20}},\ \bibinfo
  {pages} {1353} (\bibinfo {year} {2021})}\BibitemShut {NoStop}%
\bibitem [{\citenamefont {Chen}\ \emph
  {et~al.}(2021{\natexlab{a}})\citenamefont {Chen}, \citenamefont {Wang},
  \citenamefont {Yin}, \citenamefont {Gu}, \citenamefont {Jiang}, \citenamefont
  {Tu}, \citenamefont {Gong}, \citenamefont {Uwatoko}, \citenamefont {Sun},
  \citenamefont {Lei}, \citenamefont {Hu},\ and\ \citenamefont
  {Cheng}}]{chen2021double}%
  \BibitemOpen
  \bibfield  {author} {\bibinfo {author} {\bibfnamefont {K.~Y.}\ \bibnamefont
  {Chen}}, \bibinfo {author} {\bibfnamefont {N.~N.}\ \bibnamefont {Wang}},
  \bibinfo {author} {\bibfnamefont {Q.~W.}\ \bibnamefont {Yin}}, \bibinfo
  {author} {\bibfnamefont {Y.~H.}\ \bibnamefont {Gu}}, \bibinfo {author}
  {\bibfnamefont {K.}~\bibnamefont {Jiang}}, \bibinfo {author} {\bibfnamefont
  {Z.~J.}\ \bibnamefont {Tu}}, \bibinfo {author} {\bibfnamefont {C.~S.}\
  \bibnamefont {Gong}}, \bibinfo {author} {\bibfnamefont {Y.}~\bibnamefont
  {Uwatoko}}, \bibinfo {author} {\bibfnamefont {J.~P.}\ \bibnamefont {Sun}},
  \bibinfo {author} {\bibfnamefont {H.~C.}\ \bibnamefont {Lei}}, \bibinfo
  {author} {\bibfnamefont {J.~P.}\ \bibnamefont {Hu}}, and\ \bibinfo {author}
  {\bibfnamefont {J.-G.}\ \bibnamefont {Cheng}},\ }\bibfield  {title} {\bibinfo
  {title} {Double superconducting dome and triple enhancement of ${T}_{c}$ in
  the kagome superconductor {${\mathrm{CsV}}_{3}{\mathrm{Sb}}_{5}$} under high
  pressure},\ }\href {https://doi.org/10.1103/PhysRevLett.126.247001}
  {\bibfield  {journal} {\bibinfo  {journal} {Phys. Rev. Lett.}\ }\textbf
  {\bibinfo {volume} {126}},\ \bibinfo {pages} {247001} (\bibinfo {year}
  {2021}{\natexlab{a}})}\BibitemShut {NoStop}%
\bibitem [{\citenamefont {Du}\ \emph {et~al.}(2021{\natexlab{a}})\citenamefont
  {Du}, \citenamefont {Luo}, \citenamefont {Ortiz}, \citenamefont {Chen},
  \citenamefont {Duan}, \citenamefont {Zhang}, \citenamefont {Lu},
  \citenamefont {Wilson}, \citenamefont {Song},\ and\ \citenamefont
  {Yuan}}]{Du2021}%
  \BibitemOpen
  \bibfield  {author} {\bibinfo {author} {\bibfnamefont {F.}~\bibnamefont
  {Du}}, \bibinfo {author} {\bibfnamefont {S.}~\bibnamefont {Luo}}, \bibinfo
  {author} {\bibfnamefont {B.~R.}\ \bibnamefont {Ortiz}}, \bibinfo {author}
  {\bibfnamefont {Y.}~\bibnamefont {Chen}}, \bibinfo {author} {\bibfnamefont
  {W.}~\bibnamefont {Duan}}, \bibinfo {author} {\bibfnamefont {D.}~\bibnamefont
  {Zhang}}, \bibinfo {author} {\bibfnamefont {X.}~\bibnamefont {Lu}}, \bibinfo
  {author} {\bibfnamefont {S.~D.}\ \bibnamefont {Wilson}}, \bibinfo {author}
  {\bibfnamefont {Y.}~\bibnamefont {Song}}, and\ \bibinfo {author}
  {\bibfnamefont {H.}~\bibnamefont {Yuan}},\ }\bibfield  {title} {\bibinfo
  {title} {Pressure-induced double superconducting domes and charge instability
  in the kagome metal {${\mathrm{KV}}_{3}{\mathrm{Sb}}_{5}$}},\ }\href
  {https://doi.org/10.1103/PhysRevB.103.L220504} {\bibfield  {journal}
  {\bibinfo  {journal} {Phys. Rev. B}\ }\textbf {\bibinfo {volume} {103}},\
  \bibinfo {pages} {L220504} (\bibinfo {year}
  {2021}{\natexlab{a}})}\BibitemShut {NoStop}%
\bibitem [{\citenamefont {Zhang}\ \emph {et~al.}(2021)\citenamefont {Zhang},
  \citenamefont {Chen}, \citenamefont {Zhou}, \citenamefont {Yuan},
  \citenamefont {Wang}, \citenamefont {Wang}, \citenamefont {Yang},
  \citenamefont {An}, \citenamefont {Zhang}, \citenamefont {Zhu}, \citenamefont
  {Zhou}, \citenamefont {Chen}, \citenamefont {Zhou},\ and\ \citenamefont
  {Yang}}]{Zhang2021}%
  \BibitemOpen
  \bibfield  {author} {\bibinfo {author} {\bibfnamefont {Z.}~\bibnamefont
  {Zhang}}, \bibinfo {author} {\bibfnamefont {Z.}~\bibnamefont {Chen}},
  \bibinfo {author} {\bibfnamefont {Y.}~\bibnamefont {Zhou}}, \bibinfo {author}
  {\bibfnamefont {Y.}~\bibnamefont {Yuan}}, \bibinfo {author} {\bibfnamefont
  {S.}~\bibnamefont {Wang}}, \bibinfo {author} {\bibfnamefont {J.}~\bibnamefont
  {Wang}}, \bibinfo {author} {\bibfnamefont {H.}~\bibnamefont {Yang}}, \bibinfo
  {author} {\bibfnamefont {C.}~\bibnamefont {An}}, \bibinfo {author}
  {\bibfnamefont {L.}~\bibnamefont {Zhang}}, \bibinfo {author} {\bibfnamefont
  {X.}~\bibnamefont {Zhu}}, \bibinfo {author} {\bibfnamefont {Y.}~\bibnamefont
  {Zhou}}, \bibinfo {author} {\bibfnamefont {X.}~\bibnamefont {Chen}}, \bibinfo
  {author} {\bibfnamefont {J.}~\bibnamefont {Zhou}}, and\ \bibinfo {author}
  {\bibfnamefont {Z.}~\bibnamefont {Yang}},\ }\bibfield  {title} {\bibinfo
  {title} {Pressure-induced reemergence of superconductivity in the topological
  kagome metal {$\mathrm{Cs}{\mathrm{V}}_{3}{\mathrm{Sb}}_{5}$}},\ }\href
  {https://doi.org/10.1103/PhysRevB.103.224513} {\bibfield  {journal} {\bibinfo
   {journal} {Phys. Rev. B}\ }\textbf {\bibinfo {volume} {103}},\ \bibinfo
  {pages} {224513} (\bibinfo {year} {2021})}\BibitemShut {NoStop}%
\bibitem [{\citenamefont {Chen}\ \emph
  {et~al.}(2021{\natexlab{b}})\citenamefont {Chen}, \citenamefont {Zhan},
  \citenamefont {Wang}, \citenamefont {Deng}, \citenamefont {Liu},
  \citenamefont {Chen}, \citenamefont {Guo},\ and\ \citenamefont
  {Chen}}]{Chen2021}%
  \BibitemOpen
  \bibfield  {author} {\bibinfo {author} {\bibfnamefont {X.}~\bibnamefont
  {Chen}}, \bibinfo {author} {\bibfnamefont {X.}~\bibnamefont {Zhan}}, \bibinfo
  {author} {\bibfnamefont {X.}~\bibnamefont {Wang}}, \bibinfo {author}
  {\bibfnamefont {J.}~\bibnamefont {Deng}}, \bibinfo {author} {\bibfnamefont
  {X.-B.}\ \bibnamefont {Liu}}, \bibinfo {author} {\bibfnamefont
  {X.}~\bibnamefont {Chen}}, \bibinfo {author} {\bibfnamefont {J.-G.}\
  \bibnamefont {Guo}}, and\ \bibinfo {author} {\bibfnamefont {X.}~\bibnamefont
  {Chen}},\ }\bibfield  {title} {\bibinfo {title} {Highly robust reentrant
  superconductivity in {CsV$_3$Sb$_5$} under pressure},\ }\href
  {https://doi.org/10.1088/0256-307x/38/5/057402} {\bibfield  {journal}
  {\bibinfo  {journal} {Chin. Phys. Lett.}\ }\textbf {\bibinfo {volume} {38}},\
  \bibinfo {pages} {057402} (\bibinfo {year} {2021}{\natexlab{b}})}\BibitemShut
  {NoStop}%
\bibitem [{\citenamefont {Wang}\ \emph
  {et~al.}(2021{\natexlab{a}})\citenamefont {Wang}, \citenamefont {Kong},
  \citenamefont {Shi}, \citenamefont {Pei}, \citenamefont {Wen}, \citenamefont
  {Gao}, \citenamefont {Zhao}, \citenamefont {Yin}, \citenamefont {Wu},
  \citenamefont {Li}, \citenamefont {Lei}, \citenamefont {Li}, \citenamefont
  {Chen}, \citenamefont {Yan},\ and\ \citenamefont {Qi}}]{Wang2021b}%
  \BibitemOpen
  \bibfield  {author} {\bibinfo {author} {\bibfnamefont {Q.}~\bibnamefont
  {Wang}}, \bibinfo {author} {\bibfnamefont {P.}~\bibnamefont {Kong}}, \bibinfo
  {author} {\bibfnamefont {W.}~\bibnamefont {Shi}}, \bibinfo {author}
  {\bibfnamefont {C.}~\bibnamefont {Pei}}, \bibinfo {author} {\bibfnamefont
  {C.}~\bibnamefont {Wen}}, \bibinfo {author} {\bibfnamefont {L.}~\bibnamefont
  {Gao}}, \bibinfo {author} {\bibfnamefont {Y.}~\bibnamefont {Zhao}}, \bibinfo
  {author} {\bibfnamefont {Q.}~\bibnamefont {Yin}}, \bibinfo {author}
  {\bibfnamefont {Y.}~\bibnamefont {Wu}}, \bibinfo {author} {\bibfnamefont
  {G.}~\bibnamefont {Li}}, \bibinfo {author} {\bibfnamefont {H.}~\bibnamefont
  {Lei}}, \bibinfo {author} {\bibfnamefont {J.}~\bibnamefont {Li}}, \bibinfo
  {author} {\bibfnamefont {Y.}~\bibnamefont {Chen}}, \bibinfo {author}
  {\bibfnamefont {S.}~\bibnamefont {Yan}}, and\ \bibinfo {author}
  {\bibfnamefont {Y.}~\bibnamefont {Qi}},\ }\bibfield  {title} {\bibinfo
  {title} {Charge density wave orders and enhanced superconductivity under
  pressure in the kagome metal {CsV$_3$Sb$_5$}},\ }\href
  {https://doi.org/https://doi.org/10.1002/adma.202102813} {\bibfield
  {journal} {\bibinfo  {journal} {Advanced Materials}\ }\textbf {\bibinfo
  {volume} {33}},\ \bibinfo {pages} {2102813} (\bibinfo {year}
  {2021}{\natexlab{a}})}\BibitemShut {NoStop}%
\bibitem [{\citenamefont {Wang}\ \emph
  {et~al.}(2021{\natexlab{b}})\citenamefont {Wang}, \citenamefont {Chen},
  \citenamefont {Yin}, \citenamefont {Ma}, \citenamefont {Pan}, \citenamefont
  {Yang}, \citenamefont {Ji}, \citenamefont {Wu}, \citenamefont {Shan},
  \citenamefont {Xu}, \citenamefont {Tu}, \citenamefont {Gong}, \citenamefont
  {Liu}, \citenamefont {Li}, \citenamefont {Uwatoko}, \citenamefont {Dong},
  \citenamefont {Lei}, \citenamefont {Sun},\ and\ \citenamefont
  {Cheng}}]{Wang2021c}%
  \BibitemOpen
  \bibfield  {author} {\bibinfo {author} {\bibfnamefont {N.~N.}\ \bibnamefont
  {Wang}}, \bibinfo {author} {\bibfnamefont {K.~Y.}\ \bibnamefont {Chen}},
  \bibinfo {author} {\bibfnamefont {Q.~W.}\ \bibnamefont {Yin}}, \bibinfo
  {author} {\bibfnamefont {Y.~N.~N.}\ \bibnamefont {Ma}}, \bibinfo {author}
  {\bibfnamefont {B.~Y.}\ \bibnamefont {Pan}}, \bibinfo {author} {\bibfnamefont
  {X.}~\bibnamefont {Yang}}, \bibinfo {author} {\bibfnamefont {X.~Y.}\
  \bibnamefont {Ji}}, \bibinfo {author} {\bibfnamefont {S.~L.}\ \bibnamefont
  {Wu}}, \bibinfo {author} {\bibfnamefont {P.~F.}\ \bibnamefont {Shan}},
  \bibinfo {author} {\bibfnamefont {S.~X.}\ \bibnamefont {Xu}}, \bibinfo
  {author} {\bibfnamefont {Z.~J.}\ \bibnamefont {Tu}}, \bibinfo {author}
  {\bibfnamefont {C.~S.}\ \bibnamefont {Gong}}, \bibinfo {author}
  {\bibfnamefont {G.~T.}\ \bibnamefont {Liu}}, \bibinfo {author} {\bibfnamefont
  {G.}~\bibnamefont {Li}}, \bibinfo {author} {\bibfnamefont {Y.}~\bibnamefont
  {Uwatoko}}, \bibinfo {author} {\bibfnamefont {X.~L.}\ \bibnamefont {Dong}},
  \bibinfo {author} {\bibfnamefont {H.~C.}\ \bibnamefont {Lei}}, \bibinfo
  {author} {\bibfnamefont {J.~P.}\ \bibnamefont {Sun}}, and\ \bibinfo {author}
  {\bibfnamefont {J.-G.}\ \bibnamefont {Cheng}},\ }\bibfield  {title} {\bibinfo
  {title} {Competition between charge-density-wave and superconductivity in the
  kagome metal {$\mathrm{Rb}{\mathrm{V}}_{3}{\mathrm{Sb}}_{5}$}},\ }\href
  {https://doi.org/10.1103/PhysRevResearch.3.043018} {\bibfield  {journal}
  {\bibinfo  {journal} {Phys. Rev. Research}\ }\textbf {\bibinfo {volume}
  {3}},\ \bibinfo {pages} {043018} (\bibinfo {year}
  {2021}{\natexlab{b}})}\BibitemShut {NoStop}%
\bibitem [{\citenamefont {Yu}\ \emph {et~al.}(2021{\natexlab{b}})\citenamefont
  {Yu}, \citenamefont {Ma}, \citenamefont {Zhuo}, \citenamefont {Liu},
  \citenamefont {Wen}, \citenamefont {Lei}, \citenamefont {Ying},\ and\
  \citenamefont {Chen}}]{Yu2021}%
  \BibitemOpen
  \bibfield  {author} {\bibinfo {author} {\bibfnamefont {F.~H.}\ \bibnamefont
  {Yu}}, \bibinfo {author} {\bibfnamefont {D.~H.}\ \bibnamefont {Ma}}, \bibinfo
  {author} {\bibfnamefont {W.~Z.}\ \bibnamefont {Zhuo}}, \bibinfo {author}
  {\bibfnamefont {S.~Q.}\ \bibnamefont {Liu}}, \bibinfo {author} {\bibfnamefont
  {X.~K.}\ \bibnamefont {Wen}}, \bibinfo {author} {\bibfnamefont
  {B.}~\bibnamefont {Lei}}, \bibinfo {author} {\bibfnamefont {J.~J.}\
  \bibnamefont {Ying}}, and\ \bibinfo {author} {\bibfnamefont {X.~H.}\
  \bibnamefont {Chen}},\ }\bibfield  {title} {\bibinfo {title} {Unusual
  competition of superconductivity and charge-density-wave state in a
  compressed topological kagome metal},\ }\href
  {https://doi.org/10.1038/s41467-021-23928-w} {\bibfield  {journal} {\bibinfo
  {journal} {Nature Communications}\ }\textbf {\bibinfo {volume} {12}},\
  \bibinfo {pages} {3645} (\bibinfo {year} {2021}{\natexlab{b}})}\BibitemShut
  {NoStop}%
\bibitem [{\citenamefont {Du}\ \emph {et~al.}(2021{\natexlab{b}})\citenamefont
  {Du}, \citenamefont {Luo}, \citenamefont {Li}, \citenamefont {Ortiz},
  \citenamefont {Chen}, \citenamefont {Wilson}, \citenamefont {Song},\ and\
  \citenamefont {Yuan}}]{Du2021b}%
  \BibitemOpen
  \bibfield  {author} {\bibinfo {author} {\bibfnamefont {F.}~\bibnamefont
  {Du}}, \bibinfo {author} {\bibfnamefont {S.}~\bibnamefont {Luo}}, \bibinfo
  {author} {\bibfnamefont {R.}~\bibnamefont {Li}}, \bibinfo {author}
  {\bibfnamefont {B.~R.}\ \bibnamefont {Ortiz}}, \bibinfo {author}
  {\bibfnamefont {Y.}~\bibnamefont {Chen}}, \bibinfo {author} {\bibfnamefont
  {S.~D.}\ \bibnamefont {Wilson}}, \bibinfo {author} {\bibfnamefont
  {Y.}~\bibnamefont {Song}}, and\ \bibinfo {author} {\bibfnamefont
  {H.}~\bibnamefont {Yuan}},\ }\bibfield  {title} {\bibinfo {title} {Evolution
  of superconductivity and charge order in pressurized {RbV$_3$Sb$_5$}},\
  }\href {http://iopscience.iop.org/article/10.1088/1674-1056/ac4232}
  {\bibfield  {journal} {\bibinfo  {journal} {Chinese Physics B}\ } (\bibinfo
  {year} {2021}{\natexlab{b}})}\BibitemShut {NoStop}%
\bibitem [{\citenamefont {Zhao}\ \emph {et~al.}(2021)\citenamefont {Zhao},
  \citenamefont {Wang}, \citenamefont {Xia}, \citenamefont {Yin}, \citenamefont
  {Ni}, \citenamefont {Huang}, \citenamefont {Tu}, \citenamefont {Tao},
  \citenamefont {Tu}, \citenamefont {Gong}, \citenamefont {Lei}, \citenamefont
  {Guo}, \citenamefont {Yang},\ and\ \citenamefont {Li}}]{zhao2021nodal}%
  \BibitemOpen
  \bibfield  {author} {\bibinfo {author} {\bibfnamefont {C.~C.}\ \bibnamefont
  {Zhao}}, \bibinfo {author} {\bibfnamefont {L.~S.}\ \bibnamefont {Wang}},
  \bibinfo {author} {\bibfnamefont {W.}~\bibnamefont {Xia}}, \bibinfo {author}
  {\bibfnamefont {Q.~W.}\ \bibnamefont {Yin}}, \bibinfo {author} {\bibfnamefont
  {J.~M.}\ \bibnamefont {Ni}}, \bibinfo {author} {\bibfnamefont {Y.~Y.}\
  \bibnamefont {Huang}}, \bibinfo {author} {\bibfnamefont {C.~P.}\ \bibnamefont
  {Tu}}, \bibinfo {author} {\bibfnamefont {Z.~C.}\ \bibnamefont {Tao}},
  \bibinfo {author} {\bibfnamefont {Z.~J.}\ \bibnamefont {Tu}}, \bibinfo
  {author} {\bibfnamefont {C.~S.}\ \bibnamefont {Gong}}, \bibinfo {author}
  {\bibfnamefont {H.~C.}\ \bibnamefont {Lei}}, \bibinfo {author} {\bibfnamefont
  {Y.~F.}\ \bibnamefont {Guo}}, \bibinfo {author} {\bibfnamefont {X.~F.}\
  \bibnamefont {Yang}}, and\ \bibinfo {author} {\bibfnamefont {S.~Y.}\
  \bibnamefont {Li}},\ }\bibfield  {title} {\bibinfo {title} {Nodal
  superconductivity and superconducting domes in the topological kagome metal
  {CsV$_3$Sb$_5$}},\ }\href {https://arxiv.org/abs/2102.08356} {\bibfield
  {journal} {\bibinfo  {journal} {arXiv preprint arXiv:2102.08356}\ } (\bibinfo
  {year} {2021})}\BibitemShut {NoStop}%
\bibitem [{\citenamefont {Duan}\ \emph {et~al.}(2021)\citenamefont {Duan},
  \citenamefont {Nie}, \citenamefont {Luo}, \citenamefont {Yu}, \citenamefont
  {Ortiz}, \citenamefont {Yin}, \citenamefont {Su}, \citenamefont {Du},
  \citenamefont {Wang}, \citenamefont {Chen}, \citenamefont {Lu}, \citenamefont
  {Ying}, \citenamefont {Wilson}, \citenamefont {Chen}, \citenamefont {Song},\
  and\ \citenamefont {Yuan}}]{duan2021nodeless}%
  \BibitemOpen
  \bibfield  {author} {\bibinfo {author} {\bibfnamefont {W.}~\bibnamefont
  {Duan}}, \bibinfo {author} {\bibfnamefont {Z.}~\bibnamefont {Nie}}, \bibinfo
  {author} {\bibfnamefont {S.}~\bibnamefont {Luo}}, \bibinfo {author}
  {\bibfnamefont {F.}~\bibnamefont {Yu}}, \bibinfo {author} {\bibfnamefont
  {B.~R.}\ \bibnamefont {Ortiz}}, \bibinfo {author} {\bibfnamefont
  {L.}~\bibnamefont {Yin}}, \bibinfo {author} {\bibfnamefont {H.}~\bibnamefont
  {Su}}, \bibinfo {author} {\bibfnamefont {F.}~\bibnamefont {Du}}, \bibinfo
  {author} {\bibfnamefont {A.}~\bibnamefont {Wang}}, \bibinfo {author}
  {\bibfnamefont {Y.}~\bibnamefont {Chen}}, \bibinfo {author} {\bibfnamefont
  {X.}~\bibnamefont {Lu}}, \bibinfo {author} {\bibfnamefont {J.}~\bibnamefont
  {Ying}}, \bibinfo {author} {\bibfnamefont {S.~D.}\ \bibnamefont {Wilson}},
  \bibinfo {author} {\bibfnamefont {X.}~\bibnamefont {Chen}}, \bibinfo {author}
  {\bibfnamefont {Y.}~\bibnamefont {Song}}, and\ \bibinfo {author}
  {\bibfnamefont {H.}~\bibnamefont {Yuan}},\ }\bibfield  {title} {\bibinfo
  {title} {Nodeless superconductivity in the kagome metal {CsV$_3$Sb$_5$}},\
  }\href {https://doi.org/10.1007/s11433-021-1747-7} {\bibfield  {journal}
  {\bibinfo  {journal} {Science China Physics, Mechanics {\&} Astronomy}\
  }\textbf {\bibinfo {volume} {64}},\ \bibinfo {pages} {107462} (\bibinfo
  {year} {2021})}\BibitemShut {NoStop}%
\bibitem [{\citenamefont {Gupta}\ \emph {et~al.}(2021)\citenamefont {Gupta},
  \citenamefont {Das}, \citenamefont {Mielke~III}, \citenamefont {Guguchia},
  \citenamefont {Shiroka}, \citenamefont {Baines}, \citenamefont {Bartkowiak},
  \citenamefont {Luetkens}, \citenamefont {Khasanov}, \citenamefont {Yin} \emph
  {et~al.}}]{gupta2021microscopic}%
  \BibitemOpen
  \bibfield  {author} {\bibinfo {author} {\bibfnamefont {R.}~\bibnamefont
  {Gupta}}, \bibinfo {author} {\bibfnamefont {D.}~\bibnamefont {Das}}, \bibinfo
  {author} {\bibfnamefont {C.~H.}\ \bibnamefont {Mielke~III}}, \bibinfo
  {author} {\bibfnamefont {Z.}~\bibnamefont {Guguchia}}, \bibinfo {author}
  {\bibfnamefont {T.}~\bibnamefont {Shiroka}}, \bibinfo {author} {\bibfnamefont
  {C.}~\bibnamefont {Baines}}, \bibinfo {author} {\bibfnamefont
  {M.}~\bibnamefont {Bartkowiak}}, \bibinfo {author} {\bibfnamefont
  {H.}~\bibnamefont {Luetkens}}, \bibinfo {author} {\bibfnamefont
  {R.}~\bibnamefont {Khasanov}}, \bibinfo {author} {\bibfnamefont
  {Q.}~\bibnamefont {Yin}},  \emph {et~al.},\ }\bibfield  {title} {\bibinfo
  {title} {Microscopic evidence for anisotropic multigap superconductivity in
  the {CsV$_3$Sb$_5$} kagome superconductor},\ }\href
  {https://arxiv.org/abs/2108.01574} {\bibfield  {journal} {\bibinfo  {journal}
  {arXiv:2108.01574}\ } (\bibinfo {year} {2021})}\BibitemShut {NoStop}%
\bibitem [{\citenamefont {Yin}\ \emph {et~al.}(2021{\natexlab{b}})\citenamefont
  {Yin}, \citenamefont {Zhang}, \citenamefont {Chen}, \citenamefont {Ye},
  \citenamefont {Yu}, \citenamefont {Ortiz}, \citenamefont {Luo}, \citenamefont
  {Duan}, \citenamefont {Su}, \citenamefont {Ying}, \citenamefont {Wilson},
  \citenamefont {Chen}, \citenamefont {Yuan}, \citenamefont {Song},\ and\
  \citenamefont {Lu}}]{Yin2021}%
  \BibitemOpen
  \bibfield  {author} {\bibinfo {author} {\bibfnamefont {L.}~\bibnamefont
  {Yin}}, \bibinfo {author} {\bibfnamefont {D.}~\bibnamefont {Zhang}}, \bibinfo
  {author} {\bibfnamefont {C.}~\bibnamefont {Chen}}, \bibinfo {author}
  {\bibfnamefont {G.}~\bibnamefont {Ye}}, \bibinfo {author} {\bibfnamefont
  {F.}~\bibnamefont {Yu}}, \bibinfo {author} {\bibfnamefont {B.~R.}\
  \bibnamefont {Ortiz}}, \bibinfo {author} {\bibfnamefont {S.}~\bibnamefont
  {Luo}}, \bibinfo {author} {\bibfnamefont {W.}~\bibnamefont {Duan}}, \bibinfo
  {author} {\bibfnamefont {H.}~\bibnamefont {Su}}, \bibinfo {author}
  {\bibfnamefont {J.}~\bibnamefont {Ying}}, \bibinfo {author} {\bibfnamefont
  {S.~D.}\ \bibnamefont {Wilson}}, \bibinfo {author} {\bibfnamefont
  {X.}~\bibnamefont {Chen}}, \bibinfo {author} {\bibfnamefont {H.}~\bibnamefont
  {Yuan}}, \bibinfo {author} {\bibfnamefont {Y.}~\bibnamefont {Song}}, and\
  \bibinfo {author} {\bibfnamefont {X.}~\bibnamefont {Lu}},\ }\bibfield
  {title} {\bibinfo {title} {Strain-sensitive superconductivity in the kagome
  metals {${\mathrm{KV}}_{3}{\mathrm{Sb}}_{5}$ and
  $\mathrm{Cs}{\mathrm{V}}_{3}{\mathrm{Sb}}_{5}$} probed by point-contact
  spectroscopy},\ }\href {https://doi.org/10.1103/PhysRevB.104.174507}
  {\bibfield  {journal} {\bibinfo  {journal} {Phys. Rev. B}\ }\textbf {\bibinfo
  {volume} {104}},\ \bibinfo {pages} {174507} (\bibinfo {year}
  {2021}{\natexlab{b}})}\BibitemShut {NoStop}%
\bibitem [{\citenamefont {Liang}\ \emph {et~al.}(2021)\citenamefont {Liang},
  \citenamefont {Hou}, \citenamefont {Zhang}, \citenamefont {Ma}, \citenamefont
  {Wu}, \citenamefont {Zhang}, \citenamefont {Yu}, \citenamefont {Ying},
  \citenamefont {Jiang}, \citenamefont {Shan}, \citenamefont {Wang},\ and\
  \citenamefont {Chen}}]{Liang2021}%
  \BibitemOpen
  \bibfield  {author} {\bibinfo {author} {\bibfnamefont {Z.}~\bibnamefont
  {Liang}}, \bibinfo {author} {\bibfnamefont {X.}~\bibnamefont {Hou}}, \bibinfo
  {author} {\bibfnamefont {F.}~\bibnamefont {Zhang}}, \bibinfo {author}
  {\bibfnamefont {W.}~\bibnamefont {Ma}}, \bibinfo {author} {\bibfnamefont
  {P.}~\bibnamefont {Wu}}, \bibinfo {author} {\bibfnamefont {Z.}~\bibnamefont
  {Zhang}}, \bibinfo {author} {\bibfnamefont {F.}~\bibnamefont {Yu}}, \bibinfo
  {author} {\bibfnamefont {J.-J.}\ \bibnamefont {Ying}}, \bibinfo {author}
  {\bibfnamefont {K.}~\bibnamefont {Jiang}}, \bibinfo {author} {\bibfnamefont
  {L.}~\bibnamefont {Shan}}, \bibinfo {author} {\bibfnamefont {Z.}~\bibnamefont
  {Wang}}, and\ \bibinfo {author} {\bibfnamefont {X.-H.}\ \bibnamefont
  {Chen}},\ }\bibfield  {title} {\bibinfo {title} {Three-dimensional charge
  density wave and surface-dependent vortex-core states in a kagome
  superconductor {${\mathrm{CsV}}_{3}{\mathrm{Sb}}_{5}$}},\ }\href
  {https://doi.org/10.1103/PhysRevX.11.031026} {\bibfield  {journal} {\bibinfo
  {journal} {Phys. Rev. X}\ }\textbf {\bibinfo {volume} {11}},\ \bibinfo
  {pages} {031026} (\bibinfo {year} {2021})}\BibitemShut {NoStop}%
\bibitem [{\citenamefont {Xu}\ \emph {et~al.}(2021)\citenamefont {Xu},
  \citenamefont {Yan}, \citenamefont {Yin}, \citenamefont {Xia}, \citenamefont
  {Fang}, \citenamefont {Chen}, \citenamefont {Li}, \citenamefont {Yang},
  \citenamefont {Guo},\ and\ \citenamefont {Feng}}]{Xu2021}%
  \BibitemOpen
  \bibfield  {author} {\bibinfo {author} {\bibfnamefont {H.-S.}\ \bibnamefont
  {Xu}}, \bibinfo {author} {\bibfnamefont {Y.-J.}\ \bibnamefont {Yan}},
  \bibinfo {author} {\bibfnamefont {R.}~\bibnamefont {Yin}}, \bibinfo {author}
  {\bibfnamefont {W.}~\bibnamefont {Xia}}, \bibinfo {author} {\bibfnamefont
  {S.}~\bibnamefont {Fang}}, \bibinfo {author} {\bibfnamefont {Z.}~\bibnamefont
  {Chen}}, \bibinfo {author} {\bibfnamefont {Y.}~\bibnamefont {Li}}, \bibinfo
  {author} {\bibfnamefont {W.}~\bibnamefont {Yang}}, \bibinfo {author}
  {\bibfnamefont {Y.}~\bibnamefont {Guo}}, and\ \bibinfo {author}
  {\bibfnamefont {D.-L.}\ \bibnamefont {Feng}},\ }\bibfield  {title} {\bibinfo
  {title} {Multiband superconductivity with sign-preserving order parameter in
  kagome superconductor {${\mathrm{CsV}}_{3}{\mathrm{Sb}}_{5}$}},\ }\href
  {https://doi.org/10.1103/PhysRevLett.127.187004} {\bibfield  {journal}
  {\bibinfo  {journal} {Phys. Rev. Lett.}\ }\textbf {\bibinfo {volume} {127}},\
  \bibinfo {pages} {187004} (\bibinfo {year} {2021})}\BibitemShut {NoStop}%
\bibitem [{\citenamefont {Mu}\ \emph {et~al.}(2021)\citenamefont {Mu},
  \citenamefont {Yin}, \citenamefont {Tu}, \citenamefont {Gong}, \citenamefont
  {Lei}, \citenamefont {Li},\ and\ \citenamefont {Luo}}]{Mu2021}%
  \BibitemOpen
  \bibfield  {author} {\bibinfo {author} {\bibfnamefont {C.}~\bibnamefont
  {Mu}}, \bibinfo {author} {\bibfnamefont {Q.}~\bibnamefont {Yin}}, \bibinfo
  {author} {\bibfnamefont {Z.}~\bibnamefont {Tu}}, \bibinfo {author}
  {\bibfnamefont {C.}~\bibnamefont {Gong}}, \bibinfo {author} {\bibfnamefont
  {H.}~\bibnamefont {Lei}}, \bibinfo {author} {\bibfnamefont {Z.}~\bibnamefont
  {Li}}, and\ \bibinfo {author} {\bibfnamefont {J.}~\bibnamefont {Luo}},\
  }\bibfield  {title} {\bibinfo {title} {S-wave superconductivity in kagome
  metal {CsV$_3$Sb$_5$} revealed by $^{121/123}${Sb} {NQR} and {$^{51}$V} {NMR}
  measurements},\ }\href {https://doi.org/10.1088/0256-307x/38/7/077402}
  {\bibfield  {journal} {\bibinfo  {journal} {Chinese Phys. Lett.}\ }\textbf
  {\bibinfo {volume} {38}},\ \bibinfo {pages} {077402} (\bibinfo {year}
  {2021})}\BibitemShut {NoStop}%
\bibitem [{\citenamefont {Song}\ \emph {et~al.}(2021)\citenamefont {Song},
  \citenamefont {Zheng}, \citenamefont {Yu}, \citenamefont {Li}, \citenamefont
  {Nie}, \citenamefont {Shan}, \citenamefont {Zhao}, \citenamefont {Li},
  \citenamefont {Kang}, \citenamefont {Wu}, \citenamefont {Zhou}, \citenamefont
  {Sun}, \citenamefont {Liu}, \citenamefont {Luo}, \citenamefont {Wang},
  \citenamefont {Ying}, \citenamefont {Wan}, \citenamefont {Wu},\ and\
  \citenamefont {Chen}}]{song2021orbital}%
  \BibitemOpen
  \bibfield  {author} {\bibinfo {author} {\bibfnamefont {D.~W.}\ \bibnamefont
  {Song}}, \bibinfo {author} {\bibfnamefont {L.~X.}\ \bibnamefont {Zheng}},
  \bibinfo {author} {\bibfnamefont {F.~H.}\ \bibnamefont {Yu}}, \bibinfo
  {author} {\bibfnamefont {J.}~\bibnamefont {Li}}, \bibinfo {author}
  {\bibfnamefont {L.~P.}\ \bibnamefont {Nie}}, \bibinfo {author} {\bibfnamefont
  {M.}~\bibnamefont {Shan}}, \bibinfo {author} {\bibfnamefont {D.}~\bibnamefont
  {Zhao}}, \bibinfo {author} {\bibfnamefont {S.~J.}\ \bibnamefont {Li}},
  \bibinfo {author} {\bibfnamefont {B.~L.}\ \bibnamefont {Kang}}, \bibinfo
  {author} {\bibfnamefont {Z.~M.}\ \bibnamefont {Wu}}, \bibinfo {author}
  {\bibfnamefont {Y.~B.}\ \bibnamefont {Zhou}}, \bibinfo {author}
  {\bibfnamefont {K.~L.}\ \bibnamefont {Sun}}, \bibinfo {author} {\bibfnamefont
  {K.}~\bibnamefont {Liu}}, \bibinfo {author} {\bibfnamefont {X.~G.}\
  \bibnamefont {Luo}}, \bibinfo {author} {\bibfnamefont {Z.~Y.}\ \bibnamefont
  {Wang}}, \bibinfo {author} {\bibfnamefont {J.~J.}\ \bibnamefont {Ying}},
  \bibinfo {author} {\bibfnamefont {X.~G.}\ \bibnamefont {Wan}}, \bibinfo
  {author} {\bibfnamefont {T.}~\bibnamefont {Wu}}, and\ \bibinfo {author}
  {\bibfnamefont {X.~H.}\ \bibnamefont {Chen}},\ }\bibfield  {title} {\bibinfo
  {title} {Orbital ordering and fluctuations in a kagome superconductor
  {CsV$_3$Sb$_5$}},\ }\href {https://arxiv.org/abs/2104.09173} {\bibfield
  {journal} {\bibinfo  {journal} {arXiv:2104.09173}\ } (\bibinfo {year}
  {2021})}\BibitemShut {NoStop}%
\bibitem [{\citenamefont {Denner}\ \emph {et~al.}(2021)\citenamefont {Denner},
  \citenamefont {Thomale},\ and\ \citenamefont {Neupert}}]{Denner2021}%
  \BibitemOpen
  \bibfield  {author} {\bibinfo {author} {\bibfnamefont {M.~M.}\ \bibnamefont
  {Denner}}, \bibinfo {author} {\bibfnamefont {R.}~\bibnamefont {Thomale}},
  and\ \bibinfo {author} {\bibfnamefont {T.}~\bibnamefont {Neupert}},\
  }\bibfield  {title} {\bibinfo {title} {Analysis of charge order in the kagome
  metal {$A{\mathrm{V}}_{3}{\mathrm{Sb}}_{5}$
  ($A=\mathrm{K},\mathrm{Rb},\mathrm{Cs}$})},\ }\href
  {https://doi.org/10.1103/PhysRevLett.127.217601} {\bibfield  {journal}
  {\bibinfo  {journal} {Phys. Rev. Lett.}\ }\textbf {\bibinfo {volume} {127}},\
  \bibinfo {pages} {217601} (\bibinfo {year} {2021})}\BibitemShut {NoStop}%
\bibitem [{\citenamefont {Feng}\ \emph {et~al.}(2021)\citenamefont {Feng},
  \citenamefont {Jiang}, \citenamefont {Wang},\ and\ \citenamefont
  {Hu}}]{Feng2021}%
  \BibitemOpen
  \bibfield  {author} {\bibinfo {author} {\bibfnamefont {X.}~\bibnamefont
  {Feng}}, \bibinfo {author} {\bibfnamefont {K.}~\bibnamefont {Jiang}},
  \bibinfo {author} {\bibfnamefont {Z.}~\bibnamefont {Wang}}, and\ \bibinfo
  {author} {\bibfnamefont {J.}~\bibnamefont {Hu}},\ }\bibfield  {title}
  {\bibinfo {title} {Chiral flux phase in the kagome superconductor
  {$A$V$_3$Sb$_5$}},\ }\href
  {https://doi.org/https://doi.org/10.1016/j.scib.2021.04.043} {\bibfield
  {journal} {\bibinfo  {journal} {Science Bulletin}\ }\textbf {\bibinfo
  {volume} {66}},\ \bibinfo {pages} {1384} (\bibinfo {year}
  {2021})}\BibitemShut {NoStop}%
\bibitem [{\citenamefont {Tan}\ \emph {et~al.}(2021)\citenamefont {Tan},
  \citenamefont {Liu}, \citenamefont {Wang},\ and\ \citenamefont
  {Yan}}]{Tan2021}%
  \BibitemOpen
  \bibfield  {author} {\bibinfo {author} {\bibfnamefont {H.}~\bibnamefont
  {Tan}}, \bibinfo {author} {\bibfnamefont {Y.}~\bibnamefont {Liu}}, \bibinfo
  {author} {\bibfnamefont {Z.}~\bibnamefont {Wang}}, and\ \bibinfo {author}
  {\bibfnamefont {B.}~\bibnamefont {Yan}},\ }\bibfield  {title} {\bibinfo
  {title} {Charge density waves and electronic properties of superconducting
  kagome metals},\ }\href {https://doi.org/10.1103/PhysRevLett.127.046401}
  {\bibfield  {journal} {\bibinfo  {journal} {Phys. Rev. Lett.}\ }\textbf
  {\bibinfo {volume} {127}},\ \bibinfo {pages} {046401} (\bibinfo {year}
  {2021})}\BibitemShut {NoStop}%
\bibitem [{\citenamefont {Park}\ \emph {et~al.}(2021)\citenamefont {Park},
  \citenamefont {Ye},\ and\ \citenamefont {Balents}}]{Park2021}%
  \BibitemOpen
  \bibfield  {author} {\bibinfo {author} {\bibfnamefont {T.}~\bibnamefont
  {Park}}, \bibinfo {author} {\bibfnamefont {M.}~\bibnamefont {Ye}}, and\
  \bibinfo {author} {\bibfnamefont {L.}~\bibnamefont {Balents}},\ }\bibfield
  {title} {\bibinfo {title} {Electronic instabilities of kagome metals: Saddle
  points and {L}andau theory},\ }\href
  {https://doi.org/10.1103/PhysRevB.104.035142} {\bibfield  {journal} {\bibinfo
   {journal} {Phys. Rev. B}\ }\textbf {\bibinfo {volume} {104}},\ \bibinfo
  {pages} {035142} (\bibinfo {year} {2021})}\BibitemShut {NoStop}%
\bibitem [{\citenamefont {{Mielke III}}\ \emph {et~al.}(2022)\citenamefont
  {{Mielke III}}, \citenamefont {Das}, \citenamefont {Yin}, \citenamefont
  {Liu}, \citenamefont {Gupta}, \citenamefont {Wang}, \citenamefont {Jiang},
  \citenamefont {Medarde}, \citenamefont {Wu}, \citenamefont {Lei},
  \citenamefont {Chang}, \citenamefont {Dai}, \citenamefont {Si}, \citenamefont
  {Miao}, \citenamefont {Thomale}, \citenamefont {Neupert}, \citenamefont
  {Shi}, \citenamefont {Khasanov}, \citenamefont {Hasan}, \citenamefont
  {Luetkens},\ and\ \citenamefont {Guguchia}}]{Mielke2021TRS}%
  \BibitemOpen
  \bibfield  {author} {\bibinfo {author} {\bibfnamefont {C.}~\bibnamefont
  {{Mielke III}}}, \bibinfo {author} {\bibfnamefont {D.}~\bibnamefont {Das}},
  \bibinfo {author} {\bibfnamefont {J.~X.}\ \bibnamefont {Yin}}, \bibinfo
  {author} {\bibfnamefont {H.}~\bibnamefont {Liu}}, \bibinfo {author}
  {\bibfnamefont {R.}~\bibnamefont {Gupta}}, \bibinfo {author} {\bibfnamefont
  {C.~N.}\ \bibnamefont {Wang}}, \bibinfo {author} {\bibfnamefont {Y.~X.}\
  \bibnamefont {Jiang}}, \bibinfo {author} {\bibfnamefont {M.}~\bibnamefont
  {Medarde}}, \bibinfo {author} {\bibfnamefont {X.}~\bibnamefont {Wu}},
  \bibinfo {author} {\bibfnamefont {H.~C.}\ \bibnamefont {Lei}}, \bibinfo
  {author} {\bibfnamefont {J.~J.}\ \bibnamefont {Chang}}, \bibinfo {author}
  {\bibfnamefont {P.}~\bibnamefont {Dai}}, \bibinfo {author} {\bibfnamefont
  {Q.}~\bibnamefont {Si}}, \bibinfo {author} {\bibfnamefont {H.}~\bibnamefont
  {Miao}}, \bibinfo {author} {\bibfnamefont {R.}~\bibnamefont {Thomale}},
  \bibinfo {author} {\bibfnamefont {T.}~\bibnamefont {Neupert}}, \bibinfo
  {author} {\bibfnamefont {Y.}~\bibnamefont {Shi}}, \bibinfo {author}
  {\bibfnamefont {R.}~\bibnamefont {Khasanov}}, \bibinfo {author}
  {\bibfnamefont {M.~Z.}\ \bibnamefont {Hasan}}, \bibinfo {author}
  {\bibfnamefont {H.}~\bibnamefont {Luetkens}}, and\ \bibinfo {author}
  {\bibfnamefont {Z.}~\bibnamefont {Guguchia}},\ }\bibfield  {title} {\bibinfo
  {title} {Time-reversal symmetry-breaking charge order in a kagome
  superconductor},\ }\href {https://doi.org/10.1038/s41586-021-04327-z}
  {\bibfield  {journal} {\bibinfo  {journal} {Nature}\ }\textbf {\bibinfo
  {volume} {602}},\ \bibinfo {pages} {245} (\bibinfo {year}
  {2022})}\BibitemShut {NoStop}%
\bibitem [{\citenamefont {Yu}\ \emph {et~al.}(2021{\natexlab{c}})\citenamefont
  {Yu}, \citenamefont {Wang}, \citenamefont {Zhang}, \citenamefont {Sander},
  \citenamefont {Ni}, \citenamefont {Lu}, \citenamefont {Ma}, \citenamefont
  {Wang}, \citenamefont {Zhao}, \citenamefont {Chen}, \citenamefont {Jiang},
  \citenamefont {Zhang}, \citenamefont {Yang}, \citenamefont {Zhou},
  \citenamefont {Dong}, \citenamefont {Johnson}, \citenamefont {Graf},
  \citenamefont {Hu}, \citenamefont {Gao},\ and\ \citenamefont
  {Zhao}}]{yu2021evidence}%
  \BibitemOpen
  \bibfield  {author} {\bibinfo {author} {\bibfnamefont {L.}~\bibnamefont
  {Yu}}, \bibinfo {author} {\bibfnamefont {C.}~\bibnamefont {Wang}}, \bibinfo
  {author} {\bibfnamefont {Y.}~\bibnamefont {Zhang}}, \bibinfo {author}
  {\bibfnamefont {M.}~\bibnamefont {Sander}}, \bibinfo {author} {\bibfnamefont
  {S.}~\bibnamefont {Ni}}, \bibinfo {author} {\bibfnamefont {Z.}~\bibnamefont
  {Lu}}, \bibinfo {author} {\bibfnamefont {S.}~\bibnamefont {Ma}}, \bibinfo
  {author} {\bibfnamefont {Z.}~\bibnamefont {Wang}}, \bibinfo {author}
  {\bibfnamefont {Z.}~\bibnamefont {Zhao}}, \bibinfo {author} {\bibfnamefont
  {H.}~\bibnamefont {Chen}}, \bibinfo {author} {\bibfnamefont {K.}~\bibnamefont
  {Jiang}}, \bibinfo {author} {\bibfnamefont {Y.}~\bibnamefont {Zhang}},
  \bibinfo {author} {\bibfnamefont {H.}~\bibnamefont {Yang}}, \bibinfo {author}
  {\bibfnamefont {F.}~\bibnamefont {Zhou}}, \bibinfo {author} {\bibfnamefont
  {X.}~\bibnamefont {Dong}}, \bibinfo {author} {\bibfnamefont {S.~L.}\
  \bibnamefont {Johnson}}, \bibinfo {author} {\bibfnamefont {M.~J.}\
  \bibnamefont {Graf}}, \bibinfo {author} {\bibfnamefont {J.}~\bibnamefont
  {Hu}}, \bibinfo {author} {\bibfnamefont {H.-J.}\ \bibnamefont {Gao}}, and\
  \bibinfo {author} {\bibfnamefont {Z.}~\bibnamefont {Zhao}},\ }\bibfield
  {title} {\bibinfo {title} {Evidence of a hidden flux phase in the topological
  kagome metal {CsV$_3$Sb$_5$}},\ }\href {https://arxiv.org/abs/2107.10714}
  {\bibfield  {journal} {\bibinfo  {journal} {arXiv:2107.10714}\ } (\bibinfo
  {year} {2021}{\natexlab{c}})}\BibitemShut {NoStop}%
\bibitem [{\citenamefont {Li}\ \emph {et~al.}(2022)\citenamefont {Li},
  \citenamefont {Wan}, \citenamefont {Li}, \citenamefont {Li}, \citenamefont
  {Gu}, \citenamefont {Yang}, \citenamefont {Li}, \citenamefont {Wang},
  \citenamefont {Yao},\ and\ \citenamefont {Wen}}]{Li2022}%
  \BibitemOpen
  \bibfield  {author} {\bibinfo {author} {\bibfnamefont {H.}~\bibnamefont
  {Li}}, \bibinfo {author} {\bibfnamefont {S.}~\bibnamefont {Wan}}, \bibinfo
  {author} {\bibfnamefont {H.}~\bibnamefont {Li}}, \bibinfo {author}
  {\bibfnamefont {Q.}~\bibnamefont {Li}}, \bibinfo {author} {\bibfnamefont
  {Q.}~\bibnamefont {Gu}}, \bibinfo {author} {\bibfnamefont {H.}~\bibnamefont
  {Yang}}, \bibinfo {author} {\bibfnamefont {Y.}~\bibnamefont {Li}}, \bibinfo
  {author} {\bibfnamefont {Z.}~\bibnamefont {Wang}}, \bibinfo {author}
  {\bibfnamefont {Y.}~\bibnamefont {Yao}}, and\ \bibinfo {author}
  {\bibfnamefont {H.-H.}\ \bibnamefont {Wen}},\ }\bibfield  {title} {\bibinfo
  {title} {No observation of chiral flux current in the topological kagome
  metal {${\mathrm{CsV}}_{3}{\mathrm{Sb}}_{5}$}},\ }\href
  {https://doi.org/10.1103/PhysRevB.105.045102} {\bibfield  {journal} {\bibinfo
   {journal} {Phys. Rev. B}\ }\textbf {\bibinfo {volume} {105}},\ \bibinfo
  {pages} {045102} (\bibinfo {year} {2022})}\BibitemShut {NoStop}%
\bibitem [{\citenamefont {Suter}\ and\ \citenamefont
  {Wojek}(2012)}]{Suter2012}%
  \BibitemOpen
  \bibfield  {author} {\bibinfo {author} {\bibfnamefont {A.}~\bibnamefont
  {Suter}} and\ \bibinfo {author} {\bibfnamefont {B.}~\bibnamefont {Wojek}},\
  }\bibfield  {title} {\bibinfo {title} {Musrfit: {A} free platform-independent
  framework for {$\mu$SR} data analysis},\ }\href
  {https://doi.org/https://doi.org/10.1016/j.phpro.2012.04.042} {\bibfield
  {journal} {\bibinfo  {journal} {Physics Procedia}\ }\textbf {\bibinfo
  {volume} {30}},\ \bibinfo {pages} {69} (\bibinfo {year} {2012})},\ \bibinfo
  {note} {{12th International Conference on Muon Spin Rotation, Relaxation and
  Resonance ($\mu$SR2011)}}\BibitemShut {NoStop}%
\bibitem [{\citenamefont {G{\'e}ron}(2019)}]{Geron2019}%
  \BibitemOpen
  \bibfield  {author} {\bibinfo {author} {\bibfnamefont {A.}~\bibnamefont
  {G{\'e}ron}},\ }\href@noop {} {\emph {\bibinfo {title} {Hands-on machine
  learning with Scikit-Learn, Keras, and TensorFlow: Concepts, tools, and
  techniques to build intelligent systems}}}\ (\bibinfo  {publisher} {O'Reilly
  Media},\ \bibinfo {year} {2019})\BibitemShut {NoStop}%
\bibitem [{\citenamefont {Tula}\ \emph
  {et~al.}(2021{\natexlab{a}})\citenamefont {Tula}, \citenamefont {M{\"o}ller},
  \citenamefont {Quintanilla}, \citenamefont {Giblin}, \citenamefont {Hillier},
  \citenamefont {McCabe}, \citenamefont {Ramos}, \citenamefont {Barker},\ and\
  \citenamefont {Gibson}}]{Tula2021a}%
  \BibitemOpen
  \bibfield  {author} {\bibinfo {author} {\bibfnamefont {T.}~\bibnamefont
  {Tula}}, \bibinfo {author} {\bibfnamefont {G.}~\bibnamefont {M{\"o}ller}},
  \bibinfo {author} {\bibfnamefont {J.}~\bibnamefont {Quintanilla}}, \bibinfo
  {author} {\bibfnamefont {S.~R.}\ \bibnamefont {Giblin}}, \bibinfo {author}
  {\bibfnamefont {A.~D.}\ \bibnamefont {Hillier}}, \bibinfo {author}
  {\bibfnamefont {E.~E.}\ \bibnamefont {McCabe}}, \bibinfo {author}
  {\bibfnamefont {S.}~\bibnamefont {Ramos}}, \bibinfo {author} {\bibfnamefont
  {D.~S.}\ \bibnamefont {Barker}}, and\ \bibinfo {author} {\bibfnamefont
  {S.}~\bibnamefont {Gibson}},\ }\bibfield  {title} {\bibinfo {title} {Machine
  learning approach to muon spectroscopy analysis},\ }\href
  {https://doi.org/10.1088/1361-648X/abe39e} {\bibfield  {journal} {\bibinfo
  {journal} {J. Phys.: Condens. Matter}\ }\textbf {\bibinfo {volume} {33}},\
  \bibinfo {pages} {194002} (\bibinfo {year} {2021}{\natexlab{a}})}\BibitemShut
  {NoStop}%
\bibitem [{\citenamefont {Tula}\ \emph
  {et~al.}(2021{\natexlab{b}})\citenamefont {Tula}, \citenamefont {M{\"o}ller},
  \citenamefont {Quintanilla}, \citenamefont {Giblin}, \citenamefont {Hillier},
  \citenamefont {McCabe}, \citenamefont {Ramos}, \citenamefont {Barker},\ and\
  \citenamefont {Gibson}}]{Tula2021b}%
  \BibitemOpen
  \bibfield  {author} {\bibinfo {author} {\bibfnamefont {T.}~\bibnamefont
  {Tula}}, \bibinfo {author} {\bibfnamefont {G.}~\bibnamefont {M{\"o}ller}},
  \bibinfo {author} {\bibfnamefont {J.}~\bibnamefont {Quintanilla}}, \bibinfo
  {author} {\bibfnamefont {S.~R.}\ \bibnamefont {Giblin}}, \bibinfo {author}
  {\bibfnamefont {A.~D.}\ \bibnamefont {Hillier}}, \bibinfo {author}
  {\bibfnamefont {E.~E.}\ \bibnamefont {McCabe}}, \bibinfo {author}
  {\bibfnamefont {S.}~\bibnamefont {Ramos}}, \bibinfo {author} {\bibfnamefont
  {D.~S.}\ \bibnamefont {Barker}}, and\ \bibinfo {author} {\bibfnamefont
  {S.}~\bibnamefont {Gibson}},\ }\bibfield  {title} {\bibinfo {title} {Joint
  machine learning analysis of muon spectroscopy data from different
  materials},\ }\href {https://arxiv.org/abs/2112.09601} {\bibfield  {journal}
  {\bibinfo  {journal} {arXiv:2112.09601}\ } (\bibinfo {year}
  {2021}{\natexlab{b}})}\BibitemShut {NoStop}%
\bibitem [{\citenamefont {Luke}\ \emph {et~al.}(1998)\citenamefont {Luke},
  \citenamefont {Fudamoto}, \citenamefont {Kojima}, \citenamefont {Larkin},
  \citenamefont {Merrin}, \citenamefont {Nachumi}, \citenamefont {Uemura},
  \citenamefont {Maeno}, \citenamefont {Mao}, \citenamefont {Mori},
  \citenamefont {Nakamura},\ and\ \citenamefont {Sigrist}}]{Luke1998}%
  \BibitemOpen
  \bibfield  {author} {\bibinfo {author} {\bibfnamefont {G.~M.}\ \bibnamefont
  {Luke}}, \bibinfo {author} {\bibfnamefont {Y.}~\bibnamefont {Fudamoto}},
  \bibinfo {author} {\bibfnamefont {K.~M.}\ \bibnamefont {Kojima}}, \bibinfo
  {author} {\bibfnamefont {M.~I.}\ \bibnamefont {Larkin}}, \bibinfo {author}
  {\bibfnamefont {J.}~\bibnamefont {Merrin}}, \bibinfo {author} {\bibfnamefont
  {B.}~\bibnamefont {Nachumi}}, \bibinfo {author} {\bibfnamefont {Y.~J.}\
  \bibnamefont {Uemura}}, \bibinfo {author} {\bibfnamefont {Y.}~\bibnamefont
  {Maeno}}, \bibinfo {author} {\bibfnamefont {Z.~Q.}\ \bibnamefont {Mao}},
  \bibinfo {author} {\bibfnamefont {Y.}~\bibnamefont {Mori}}, \bibinfo {author}
  {\bibfnamefont {H.}~\bibnamefont {Nakamura}}, and\ \bibinfo {author}
  {\bibfnamefont {M.}~\bibnamefont {Sigrist}},\ }\bibfield  {title} {\bibinfo
  {title} {Time-reversal symmetry-breaking superconductivity in
  {${\mathrm{Sr}}_{2}{\mathrm{RuO}}_{4}$}},\ }\href
  {https://doi.org/10.1038/29038} {\bibfield  {journal} {\bibinfo  {journal}
  {Nature}\ }\textbf {\bibinfo {volume} {394}},\ \bibinfo {pages} {558}
  (\bibinfo {year} {1998})}\BibitemShut {NoStop}%
\bibitem [{\citenamefont {Hillier}\ \emph {et~al.}(2009)\citenamefont
  {Hillier}, \citenamefont {Quintanilla},\ and\ \citenamefont
  {Cywinski}}]{Hillier2009}%
  \BibitemOpen
  \bibfield  {author} {\bibinfo {author} {\bibfnamefont {A.~D.}\ \bibnamefont
  {Hillier}}, \bibinfo {author} {\bibfnamefont {J.}~\bibnamefont
  {Quintanilla}}, and\ \bibinfo {author} {\bibfnamefont {R.}~\bibnamefont
  {Cywinski}},\ }\bibfield  {title} {\bibinfo {title} {Evidence for
  time-reversal symmetry breaking in the noncentrosymmetric superconductor
  {${\mathrm{LaNiC}}_{2}$}},\ }\href
  {https://doi.org/10.1103/PhysRevLett.102.117007} {\bibfield  {journal}
  {\bibinfo  {journal} {Phys. Rev. Lett.}\ }\textbf {\bibinfo {volume} {102}},\
  \bibinfo {pages} {117007} (\bibinfo {year} {2009})}\BibitemShut {NoStop}%
\bibitem [{\citenamefont {Grinenko}\ \emph {et~al.}(2020)\citenamefont
  {Grinenko}, \citenamefont {Sarkar}, \citenamefont {Kihou}, \citenamefont
  {Lee}, \citenamefont {Morozov}, \citenamefont {Aswartham}, \citenamefont
  {B{\"u}chner}, \citenamefont {Chekhonin}, \citenamefont {Skrotzki},
  \citenamefont {Nenkov}, \citenamefont {H{\"u}hne}, \citenamefont {Nielsch},
  \citenamefont {Drechsler}, \citenamefont {Vadimov}, \citenamefont {Silaev},
  \citenamefont {Volkov}, \citenamefont {Eremin}, \citenamefont {Luetkens},\
  and\ \citenamefont {Klauss}}]{Grinenko2020}%
  \BibitemOpen
  \bibfield  {author} {\bibinfo {author} {\bibfnamefont {V.}~\bibnamefont
  {Grinenko}}, \bibinfo {author} {\bibfnamefont {R.}~\bibnamefont {Sarkar}},
  \bibinfo {author} {\bibfnamefont {K.}~\bibnamefont {Kihou}}, \bibinfo
  {author} {\bibfnamefont {C.~H.}\ \bibnamefont {Lee}}, \bibinfo {author}
  {\bibfnamefont {I.}~\bibnamefont {Morozov}}, \bibinfo {author} {\bibfnamefont
  {S.}~\bibnamefont {Aswartham}}, \bibinfo {author} {\bibfnamefont
  {B.}~\bibnamefont {B{\"u}chner}}, \bibinfo {author} {\bibfnamefont
  {P.}~\bibnamefont {Chekhonin}}, \bibinfo {author} {\bibfnamefont
  {W.}~\bibnamefont {Skrotzki}}, \bibinfo {author} {\bibfnamefont
  {K.}~\bibnamefont {Nenkov}}, \bibinfo {author} {\bibfnamefont
  {R.}~\bibnamefont {H{\"u}hne}}, \bibinfo {author} {\bibfnamefont
  {K.}~\bibnamefont {Nielsch}}, \bibinfo {author} {\bibfnamefont {S.~L.}\
  \bibnamefont {Drechsler}}, \bibinfo {author} {\bibfnamefont {V.~L.}\
  \bibnamefont {Vadimov}}, \bibinfo {author} {\bibfnamefont {M.~A.}\
  \bibnamefont {Silaev}}, \bibinfo {author} {\bibfnamefont {P.~A.}\
  \bibnamefont {Volkov}}, \bibinfo {author} {\bibfnamefont {I.}~\bibnamefont
  {Eremin}}, \bibinfo {author} {\bibfnamefont {H.}~\bibnamefont {Luetkens}},
  and\ \bibinfo {author} {\bibfnamefont {H.-H.}\ \bibnamefont {Klauss}},\
  }\bibfield  {title} {\bibinfo {title} {Superconductivity with broken
  time-reversal symmetry inside a superconducting $s$-wave state},\ }\href
  {https://doi.org/10.1038/s41567-020-0886-9} {\bibfield  {journal} {\bibinfo
  {journal} {Nature Physics}\ }\textbf {\bibinfo {volume} {16}},\ \bibinfo
  {pages} {789} (\bibinfo {year} {2020})}\BibitemShut {NoStop}%
\bibitem [{\citenamefont {Biswas}\ \emph {et~al.}(2013)\citenamefont {Biswas},
  \citenamefont {Luetkens}, \citenamefont {Neupert}, \citenamefont {St\"urzer},
  \citenamefont {Baines}, \citenamefont {Pascua}, \citenamefont {Schnyder},
  \citenamefont {Fischer}, \citenamefont {Goryo}, \citenamefont {Lees},
  \citenamefont {Maeter}, \citenamefont {Br\"uckner}, \citenamefont {Klauss},
  \citenamefont {Nicklas}, \citenamefont {Baker}, \citenamefont {Hillier},
  \citenamefont {Sigrist}, \citenamefont {Amato},\ and\ \citenamefont
  {Johrendt}}]{Biswas2013}%
  \BibitemOpen
  \bibfield  {author} {\bibinfo {author} {\bibfnamefont {P.~K.}\ \bibnamefont
  {Biswas}}, \bibinfo {author} {\bibfnamefont {H.}~\bibnamefont {Luetkens}},
  \bibinfo {author} {\bibfnamefont {T.}~\bibnamefont {Neupert}}, \bibinfo
  {author} {\bibfnamefont {T.}~\bibnamefont {St\"urzer}}, \bibinfo {author}
  {\bibfnamefont {C.}~\bibnamefont {Baines}}, \bibinfo {author} {\bibfnamefont
  {G.}~\bibnamefont {Pascua}}, \bibinfo {author} {\bibfnamefont {A.~P.}\
  \bibnamefont {Schnyder}}, \bibinfo {author} {\bibfnamefont {M.~H.}\
  \bibnamefont {Fischer}}, \bibinfo {author} {\bibfnamefont {J.}~\bibnamefont
  {Goryo}}, \bibinfo {author} {\bibfnamefont {M.~R.}\ \bibnamefont {Lees}},
  \bibinfo {author} {\bibfnamefont {H.}~\bibnamefont {Maeter}}, \bibinfo
  {author} {\bibfnamefont {F.}~\bibnamefont {Br\"uckner}}, \bibinfo {author}
  {\bibfnamefont {H.-H.}\ \bibnamefont {Klauss}}, \bibinfo {author}
  {\bibfnamefont {M.}~\bibnamefont {Nicklas}}, \bibinfo {author} {\bibfnamefont
  {P.~J.}\ \bibnamefont {Baker}}, \bibinfo {author} {\bibfnamefont {A.~D.}\
  \bibnamefont {Hillier}}, \bibinfo {author} {\bibfnamefont {M.}~\bibnamefont
  {Sigrist}}, \bibinfo {author} {\bibfnamefont {A.}~\bibnamefont {Amato}}, and\
  \bibinfo {author} {\bibfnamefont {D.}~\bibnamefont {Johrendt}},\ }\bibfield
  {title} {\bibinfo {title} {Evidence for superconductivity with broken
  time-reversal symmetry in locally noncentrosymmetric {SrPtAs}},\ }\href
  {https://doi.org/10.1103/PhysRevB.87.180503} {\bibfield  {journal} {\bibinfo
  {journal} {Phys. Rev. B}\ }\textbf {\bibinfo {volume} {87}},\ \bibinfo
  {pages} {180503} (\bibinfo {year} {2013})}\BibitemShut {NoStop}%
\bibitem [{\citenamefont {Shang}\ \emph {et~al.}(2020)\citenamefont {Shang},
  \citenamefont {Smidman}, \citenamefont {Wang}, \citenamefont {Chang},
  \citenamefont {Baines}, \citenamefont {Lee}, \citenamefont {Nie},
  \citenamefont {Pang}, \citenamefont {Xie}, \citenamefont {Jiang},
  \citenamefont {Shi}, \citenamefont {Medarde}, \citenamefont {Shiroka},\ and\
  \citenamefont {Yuan}}]{Shang2020}%
  \BibitemOpen
  \bibfield  {author} {\bibinfo {author} {\bibfnamefont {T.}~\bibnamefont
  {Shang}}, \bibinfo {author} {\bibfnamefont {M.}~\bibnamefont {Smidman}},
  \bibinfo {author} {\bibfnamefont {A.}~\bibnamefont {Wang}}, \bibinfo {author}
  {\bibfnamefont {L.-J.}\ \bibnamefont {Chang}}, \bibinfo {author}
  {\bibfnamefont {C.}~\bibnamefont {Baines}}, \bibinfo {author} {\bibfnamefont
  {M.~K.}\ \bibnamefont {Lee}}, \bibinfo {author} {\bibfnamefont {Z.~Y.}\
  \bibnamefont {Nie}}, \bibinfo {author} {\bibfnamefont {G.~M.}\ \bibnamefont
  {Pang}}, \bibinfo {author} {\bibfnamefont {W.}~\bibnamefont {Xie}}, \bibinfo
  {author} {\bibfnamefont {W.~B.}\ \bibnamefont {Jiang}}, \bibinfo {author}
  {\bibfnamefont {M.}~\bibnamefont {Shi}}, \bibinfo {author} {\bibfnamefont
  {M.}~\bibnamefont {Medarde}}, \bibinfo {author} {\bibfnamefont
  {T.}~\bibnamefont {Shiroka}}, and\ \bibinfo {author} {\bibfnamefont {H.~Q.}\
  \bibnamefont {Yuan}},\ }\bibfield  {title} {\bibinfo {title} {Simultaneous
  nodal superconductivity and time-reversal symmetry breaking in the
  noncentrosymmetric superconductor {CaPtAs}},\ }\href
  {https://doi.org/10.1103/PhysRevLett.124.207001} {\bibfield  {journal}
  {\bibinfo  {journal} {Phys. Rev. Lett.}\ }\textbf {\bibinfo {volume} {124}},\
  \bibinfo {pages} {207001} (\bibinfo {year} {2020})}\BibitemShut {NoStop}%
\bibitem [{\citenamefont {Kenney}\ \emph {et~al.}(2021)\citenamefont {Kenney},
  \citenamefont {Ortiz}, \citenamefont {Wang}, \citenamefont {Wilson},\ and\
  \citenamefont {Graf}}]{kenney2020absence}%
  \BibitemOpen
  \bibfield  {author} {\bibinfo {author} {\bibfnamefont {E.~M.}\ \bibnamefont
  {Kenney}}, \bibinfo {author} {\bibfnamefont {B.~R.}\ \bibnamefont {Ortiz}},
  \bibinfo {author} {\bibfnamefont {C.}~\bibnamefont {Wang}}, \bibinfo {author}
  {\bibfnamefont {S.~D.}\ \bibnamefont {Wilson}}, and\ \bibinfo {author}
  {\bibfnamefont {M.~J.}\ \bibnamefont {Graf}},\ }\bibfield  {title} {\bibinfo
  {title} {Absence of local moments in the kagome metal {KV}$_3${S}b$_5$ as
  determined by muon spin spectroscopy},\ }\href
  {https://doi.org/10.1088/1361-648x/abe8f9} {\bibfield  {journal} {\bibinfo
  {journal} {J. Phys.: Condens. Matter}\ }\textbf {\bibinfo {volume} {33}},\
  \bibinfo {pages} {235801} (\bibinfo {year} {2021})}\BibitemShut {NoStop}%
\bibitem [{\citenamefont {Aoki}\ \emph {et~al.}(2003)\citenamefont {Aoki},
  \citenamefont {Tsuchiya}, \citenamefont {Kanayama}, \citenamefont {Saha},
  \citenamefont {Sugawara}, \citenamefont {Sato}, \citenamefont {Higemoto},
  \citenamefont {Koda}, \citenamefont {Ohishi}, \citenamefont {Nishiyama},\
  and\ \citenamefont {Kadono}}]{Aoki2003}%
  \BibitemOpen
  \bibfield  {author} {\bibinfo {author} {\bibfnamefont {Y.}~\bibnamefont
  {Aoki}}, \bibinfo {author} {\bibfnamefont {A.}~\bibnamefont {Tsuchiya}},
  \bibinfo {author} {\bibfnamefont {T.}~\bibnamefont {Kanayama}}, \bibinfo
  {author} {\bibfnamefont {S.~R.}\ \bibnamefont {Saha}}, \bibinfo {author}
  {\bibfnamefont {H.}~\bibnamefont {Sugawara}}, \bibinfo {author}
  {\bibfnamefont {H.}~\bibnamefont {Sato}}, \bibinfo {author} {\bibfnamefont
  {W.}~\bibnamefont {Higemoto}}, \bibinfo {author} {\bibfnamefont
  {A.}~\bibnamefont {Koda}}, \bibinfo {author} {\bibfnamefont {K.}~\bibnamefont
  {Ohishi}}, \bibinfo {author} {\bibfnamefont {K.}~\bibnamefont {Nishiyama}},
  and\ \bibinfo {author} {\bibfnamefont {R.}~\bibnamefont {Kadono}},\
  }\bibfield  {title} {\bibinfo {title} {Time-reversal symmetry-breaking
  superconductivity in heavy-fermion
  {${\mathrm{P}\mathrm{r}\mathrm{O}\mathrm{s}}_{4}{\mathrm{S}\mathrm{b}}_{12}$}
  detected by muon-spin relaxation},\ }\href
  {https://doi.org/10.1103/PhysRevLett.91.067003} {\bibfield  {journal}
  {\bibinfo  {journal} {Phys. Rev. Lett.}\ }\textbf {\bibinfo {volume} {91}},\
  \bibinfo {pages} {067003} (\bibinfo {year} {2003})}\BibitemShut {NoStop}%
\bibitem [{\citenamefont {Ratcliff}\ \emph {et~al.}(2021)\citenamefont
  {Ratcliff}, \citenamefont {Hallett}, \citenamefont {Ortiz}, \citenamefont
  {Wilson},\ and\ \citenamefont {Harter}}]{Ratcliff2021}%
  \BibitemOpen
  \bibfield  {author} {\bibinfo {author} {\bibfnamefont {N.}~\bibnamefont
  {Ratcliff}}, \bibinfo {author} {\bibfnamefont {L.}~\bibnamefont {Hallett}},
  \bibinfo {author} {\bibfnamefont {B.~R.}\ \bibnamefont {Ortiz}}, \bibinfo
  {author} {\bibfnamefont {S.~D.}\ \bibnamefont {Wilson}}, and\ \bibinfo
  {author} {\bibfnamefont {J.~W.}\ \bibnamefont {Harter}},\ }\bibfield  {title}
  {\bibinfo {title} {Coherent phonon spectroscopy and interlayer modulation of
  charge density wave order in the kagome metal
  {${\mathrm{CsV}}_{3}{\mathrm{Sb}}_{5}$}},\ }\href
  {https://doi.org/10.1103/PhysRevMaterials.5.L111801} {\bibfield  {journal}
  {\bibinfo  {journal} {Phys. Rev. Materials}\ }\textbf {\bibinfo {volume}
  {5}},\ \bibinfo {pages} {L111801} (\bibinfo {year} {2021})}\BibitemShut
  {NoStop}%
\bibitem [{\citenamefont {Wang}\ \emph
  {et~al.}(2021{\natexlab{c}})\citenamefont {Wang}, \citenamefont {Wu},
  \citenamefont {Yin}, \citenamefont {Gong}, \citenamefont {Tu}, \citenamefont
  {Lin}, \citenamefont {Liu}, \citenamefont {Shi}, \citenamefont {Zhang},
  \citenamefont {Wu}, \citenamefont {Lei}, \citenamefont {Dong},\ and\
  \citenamefont {Wang}}]{Wang2021}%
  \BibitemOpen
  \bibfield  {author} {\bibinfo {author} {\bibfnamefont {Z.~X.}\ \bibnamefont
  {Wang}}, \bibinfo {author} {\bibfnamefont {Q.}~\bibnamefont {Wu}}, \bibinfo
  {author} {\bibfnamefont {Q.~W.}\ \bibnamefont {Yin}}, \bibinfo {author}
  {\bibfnamefont {C.~S.}\ \bibnamefont {Gong}}, \bibinfo {author}
  {\bibfnamefont {Z.~J.}\ \bibnamefont {Tu}}, \bibinfo {author} {\bibfnamefont
  {T.}~\bibnamefont {Lin}}, \bibinfo {author} {\bibfnamefont {Q.~M.}\
  \bibnamefont {Liu}}, \bibinfo {author} {\bibfnamefont {L.~Y.}\ \bibnamefont
  {Shi}}, \bibinfo {author} {\bibfnamefont {S.~J.}\ \bibnamefont {Zhang}},
  \bibinfo {author} {\bibfnamefont {D.}~\bibnamefont {Wu}}, \bibinfo {author}
  {\bibfnamefont {H.~C.}\ \bibnamefont {Lei}}, \bibinfo {author} {\bibfnamefont
  {T.}~\bibnamefont {Dong}}, and\ \bibinfo {author} {\bibfnamefont {N.~L.}\
  \bibnamefont {Wang}},\ }\bibfield  {title} {\bibinfo {title} {Unconventional
  charge density wave and photoinduced lattice symmetry change in the kagome
  metal {${\mathrm{CsV}}_{3}{\mathrm{Sb}}_{5}$} probed by time-resolved
  spectroscopy},\ }\href {https://doi.org/10.1103/PhysRevB.104.165110}
  {\bibfield  {journal} {\bibinfo  {journal} {Phys. Rev. B}\ }\textbf {\bibinfo
  {volume} {104}},\ \bibinfo {pages} {165110} (\bibinfo {year}
  {2021}{\natexlab{c}})}\BibitemShut {NoStop}%
\bibitem [{\citenamefont {Ma{\~n}as-Valero}\ \emph {et~al.}(2021)\citenamefont
  {Ma{\~n}as-Valero}, \citenamefont {Huddart}, \citenamefont {Lancaster},
  \citenamefont {Coronado},\ and\ \citenamefont {Pratt}}]{manas2021quantum}%
  \BibitemOpen
  \bibfield  {author} {\bibinfo {author} {\bibfnamefont {S.}~\bibnamefont
  {Ma{\~n}as-Valero}}, \bibinfo {author} {\bibfnamefont {B.~M.}\ \bibnamefont
  {Huddart}}, \bibinfo {author} {\bibfnamefont {T.}~\bibnamefont {Lancaster}},
  \bibinfo {author} {\bibfnamefont {E.}~\bibnamefont {Coronado}}, and\ \bibinfo
  {author} {\bibfnamefont {F.~L.}\ \bibnamefont {Pratt}},\ }\bibfield  {title}
  {\bibinfo {title} {Quantum phases and spin liquid properties of
  {1T-TaS$_2$}},\ }\href {https://doi.org/10.1038/s41535-021-00367-w}
  {\bibfield  {journal} {\bibinfo  {journal} {npj Quantum Materials}\ }\textbf
  {\bibinfo {volume} {6}},\ \bibinfo {pages} {1} (\bibinfo {year}
  {2021})}\BibitemShut {NoStop}%
\bibitem [{\citenamefont {Lin}\ and\ \citenamefont
  {Nandkishore}()}]{lin2021kagome}%
  \BibitemOpen
  \bibfield  {author} {\bibinfo {author} {\bibfnamefont {Y.-P.}\ \bibnamefont
  {Lin}} and\ \bibinfo {author} {\bibfnamefont {R.~M.}\ \bibnamefont
  {Nandkishore}},\ }\bibfield  {title} {\bibinfo {title} {Kagome
  superconductors from {P}omeranchuk fluctuations in charge density wave
  metals},\ }\href {https://arxiv.org/abs/2107.09050} {\bibinfo  {journal}
  {arXiv:2107.09050}\ }\BibitemShut {NoStop}%
\bibitem [{\citenamefont {Ghosh}\ \emph {et~al.}(2020)\citenamefont {Ghosh},
  \citenamefont {Smidman}, \citenamefont {Shang}, \citenamefont {Annett},
  \citenamefont {Hillier}, \citenamefont {Quintanilla},\ and\ \citenamefont
  {Yuan}}]{Ghosh2020}%
  \BibitemOpen
\bibfield  {journal} {  }\bibfield  {author} {\bibinfo {author} {\bibfnamefont
  {S.~K.}\ \bibnamefont {Ghosh}}, \bibinfo {author} {\bibfnamefont
  {M.}~\bibnamefont {Smidman}}, \bibinfo {author} {\bibfnamefont
  {T.}~\bibnamefont {Shang}}, \bibinfo {author} {\bibfnamefont {J.~F.}\
  \bibnamefont {Annett}}, \bibinfo {author} {\bibfnamefont {A.~D.}\
  \bibnamefont {Hillier}}, \bibinfo {author} {\bibfnamefont {J.}~\bibnamefont
  {Quintanilla}}, and\ \bibinfo {author} {\bibfnamefont {H.}~\bibnamefont
  {Yuan}},\ }\bibfield  {title} {\bibinfo {title} {Recent progress on
  superconductors with time-reversal symmetry breaking},\ }\href
  {https://doi.org/10.1088/1361-648x/abaa06} {\bibfield  {journal} {\bibinfo
  {journal} {J. Phys.: Condens. Matter}\ }\textbf {\bibinfo {volume} {33}},\
  \bibinfo {pages} {033001} (\bibinfo {year} {2020})}\BibitemShut {NoStop}%
\bibitem [{\citenamefont {Wu}\ \emph {et~al.}(2021)\citenamefont {Wu},
  \citenamefont {Schwemmer}, \citenamefont {M\"uller}, \citenamefont
  {Consiglio}, \citenamefont {Sangiovanni}, \citenamefont {Di~Sante},
  \citenamefont {Iqbal}, \citenamefont {Hanke}, \citenamefont {Schnyder},
  \citenamefont {Denner}, \citenamefont {Fischer}, \citenamefont {Neupert},\
  and\ \citenamefont {Thomale}}]{wu2021nature}%
  \BibitemOpen
  \bibfield  {author} {\bibinfo {author} {\bibfnamefont {X.}~\bibnamefont
  {Wu}}, \bibinfo {author} {\bibfnamefont {T.}~\bibnamefont {Schwemmer}},
  \bibinfo {author} {\bibfnamefont {T.}~\bibnamefont {M\"uller}}, \bibinfo
  {author} {\bibfnamefont {A.}~\bibnamefont {Consiglio}}, \bibinfo {author}
  {\bibfnamefont {G.}~\bibnamefont {Sangiovanni}}, \bibinfo {author}
  {\bibfnamefont {D.}~\bibnamefont {Di~Sante}}, \bibinfo {author}
  {\bibfnamefont {Y.}~\bibnamefont {Iqbal}}, \bibinfo {author} {\bibfnamefont
  {W.}~\bibnamefont {Hanke}}, \bibinfo {author} {\bibfnamefont {A.~P.}\
  \bibnamefont {Schnyder}}, \bibinfo {author} {\bibfnamefont {M.~M.}\
  \bibnamefont {Denner}}, \bibinfo {author} {\bibfnamefont {M.~H.}\
  \bibnamefont {Fischer}}, \bibinfo {author} {\bibfnamefont {T.}~\bibnamefont
  {Neupert}}, and\ \bibinfo {author} {\bibfnamefont {R.}~\bibnamefont
  {Thomale}},\ }\bibfield  {title} {\bibinfo {title} {Nature of unconventional
  pairing in the kagome superconductors {$A{\mathrm{V}}_{3}{\mathrm{Sb}}_{5}$
  ($A=\mathrm{K},\mathrm{Rb},\mathrm{Cs}$)}},\ }\href
  {https://doi.org/10.1103/PhysRevLett.127.177001} {\bibfield  {journal}
  {\bibinfo  {journal} {Phys. Rev. Lett.}\ }\textbf {\bibinfo {volume} {127}},\
  \bibinfo {pages} {177001} (\bibinfo {year} {2021})}\BibitemShut {NoStop}%
\bibitem [{RB1()}]{RB1}%
  \BibitemOpen
  \href@noop {} {}\bibinfo {note} {{M. Smidman et al; (2021): STFC ISIS Neutron
  and Muon Source, \url{https://doi.org/10.5286/ISIS.E.RB2000246}}}\BibitemShut
  {NoStop}%
\bibitem [{RB2()}]{RB2}%
  \BibitemOpen
  \href@noop {} {}\bibinfo {note} {{S. K. Ghosh et al; (2021): STFC ISIS
  Neutron and Muon Source, \url{https://doi.org/10.5286/ISIS.E.RB2000245
  }}}\BibitemShut {NoStop}%
\end{thebibliography}

%

\end{document}